\documentclass{emulateapj}
\usepackage{aas_macros}
\usepackage{amssymb}
\usepackage{amsmath}

\usepackage{natbib}
\usepackage[svgnames]{xcolor}
\makeatletter

\newcommand{\Rmnum}[1]{\expandafter\@slowromancap\romannumeral #1@}
\makeatother

\newcommand  \HII{\,H\,{\footnotesize II}}

\newcommand  \m{\mathrm}

\begin{document}
\title{What sets the massive star formation rates and efficiencies \\ of giant molecular clouds?}
\author{Bram B. Ochsendorf\altaffilmark{1}, Margaret Meixner\altaffilmark{1,2}, Julia Roman-Duval\altaffilmark{2}, Mubdi Rahman\altaffilmark{1} \& Neal J. Evans II\altaffilmark{3,4}}
\affil{$^1$ Department of Physics and Astronomy, The Johns Hopkins University, 3400 North Charles Street, Baltimore, MD 21218, USA\\ $^2$ Space Telescope Science Institute, 3700 San Martin Drive, Baltimore, MD 21218, USA \\ $^3$ Department of Astronomy, The University of Texas at Austin, 2515 Speedway, Stop C1400, Austin, TX 78712-1205, USA \\ $^4$ Korea Astronomy and Space Science Institute, 776 Daedeokdae-ro, Yuseong-gu, Daejeon, 34055, Korea}
\email{bochsen1@jhu.edu}

\begin{abstract}
Galactic star formation scaling relations show increased scatter from kpc to sub-kpc scales. Investigating this scatter may hold important clues to how the star formation process evolves in time and space. Here, we combine different molecular gas tracers, different star formation indicators probing distinct populations of massive stars, and knowledge on the evolutionary state of each star forming region to derive star formation properties of $\sim$\,150 star forming complexes over the face of the Large Magellanic Cloud. We find that the rate of massive star formation ramps up when stellar clusters emerge and boost the formation of subsequent generations of massive stars. In addition, we reveal that the star formation efficiency of individual GMCs declines with increasing cloud gas mass ($M_\m{cloud}$). This trend persists in Galactic star forming regions, and implies higher molecular gas depletion times for larger GMCs. 

We compare the star formation efficiency per freefall time ($\epsilon_\m{ff}$) with predictions from various widely-used analytical star formation models. We show that while these models can produce large dispersions in $\epsilon_\m{ff}$ similar to observations, the origin of the model-predicted scatter is inconsistent with observations. Moreover, all models fail to reproduce the observed decline of  $\epsilon_\m{ff}$ with increasing $M_\m{cloud}$ in the LMC and the Milky Way. We conclude that analytical star formation models idealizing global turbulence levels, cloud densities, and assuming a stationary SFR are inconsistent with observations from modern datasets tracing massive star formation on individual cloud scales. Instead, we reiterate the importance of local stellar feedback in shaping the properties of GMCs and setting their massive star formation rate.

\end{abstract}

\section{Introduction}

Star formation studies have found relations between molecular gas and star formation rate (SFR) on kpc scales \citep[see][and references thererin]{kennicutt_2012}. With the advent of high-resolution, sensitive ground and space-based observatories, we are able to test these relations on smaller scales. In this respect, it has been shown that star formation efficiencies vary over 2\,-\,3 orders of magnitudes between individual star forming regions when star formation is measured using the emission from massive stars \citep{mooney_1988, mead_1990, vutisalchavakul_2016, lee_2016}. This raises the question: what sets the massive star formation rate of giant molecular clouds (GMCs)?

Massive stars likely form within massive, dense {\em cores} \citep{tan_2014}, which are substructures within more massive, dense {\em clumps}, the birthplace of clusters \citep{mckee_2007}. Thus, a viable explanation of the large scatter in star formation efficiency between individual GMCs is a varying dense gas fraction \citep{lada_2012,krumholz_2012}. Indeed, the SFR within GMCs appears to be linearly correlated with dense gas \citep{wu_2005}, indicating that the internal structure of GMCs plays a pivotal role in setting the global (massive) SFR. It is therefore of paramount interest to study the evolution of GMCs, identify the physical mechanisms that regulate their structure and, consequently, the (massive) star formation rate.

There is growing consensus that GMCs are supported by supersonic turbulence \citep{mac-low_2004,dobbs_2014}. In this scenario, star formation occurs in the fraction of GMCs containing gas that overcomes this support and collapses under self-gravity. Various analytic `turbulence-regulated' star formation models have been proposed in recent years \citep[e.g.,][]{krumholz_2005,hennebelle_2011,padoan_2011,federrath_2012} that attempt to explain the observed star formation properties of GMCs. However, there is an outstanding lack of comparison with observations, which mostly rely on those of nearby molecular clouds \citep{evans_2009,lada_2010,heiderman_2010,evans_2014} and/or a relatively small sample of dense gas clumps in the Galaxy \citep{wu_2005}. In this respect, recent works that employ larger samples of Galactic GMCs \citep{vutisalchavakul_2016,lee_2016} appear to challenge theory by showing large scatter in star formation efficiencies that are not reproduced by the aforementioned star formation models.

The Large Magellanic Cloud (LMC), with its proximity and favorable orientation, limits confusion and places clouds and stars at a common known distance, which allows for detailed studies of massive star formation on a galaxy-wide scale. In \citet{ochsendorf_2016}, we determined that massive young stellar objects (MYSOs) in the LMC are not typically found at the highest column densities nor centers of their parent GMCs on $\sim$\,5 pc scales, while their number density is significantly boosted near young ($\textless$\,10 Myr) stellar clusters. These results reveal a ubiquitous connection between different generations of massive stars on timescales up to 10 Myr, and may illustrate the importance of stellar feedback in shaping how gas collapses under self-gravity. Here, we build upon these results by deriving SFRs and star formation efficiencies of $\sim$\,150 individual star forming complexes over the face of the Large Magellanic Cloud, and comparing these with widely-used star formation models. Our method distinguishes itself from the Galactic studies of \citet{vutisalchavakul_2016} and \citet{lee_2016} by (1) using different SFR tracers probing distinct populations of massive stars, (2) employing previous knowledge on the evolutionary state of each star forming region, (3) testing the dependency of the results on the mass and structure of GMCs with different molecular gas tracers, and (4) determining if the derived relations hold from the Milky Way to the ISM in a low-metallicity dwarf galaxy. The observations and method are described in Sec. \ref{sec:method}, results of the analysis in Sec. \ref{sec:results}, and we discuss the implications in Sec. \ref{sec:discussion}. We summarize our main conclusions in Sec. \ref{sec:conclusions}.

\section{Method}\label{sec:method}

There are a variety of ways to separate emission into clouds and to measure their masses. We employ two methods: one uses the traditional conversion from CO luminosity to mass (denoted MAGMA); the second uses the dust continuum emission (denoted J16). Similarly, we use two methods to measure the star formation rate: one counts massive YSOs and uses an IMF and a characteristic age, as employed in studies of nearby Galactic clouds (denoted MYSOs); the other uses the diffuse emission from gas and dust affected by star formation, as is traditional in extragalactic studies (denoted H$\alpha$ + 24 $\mu$m).

\begin{figure*}
\centering
\includegraphics[width=17cm]{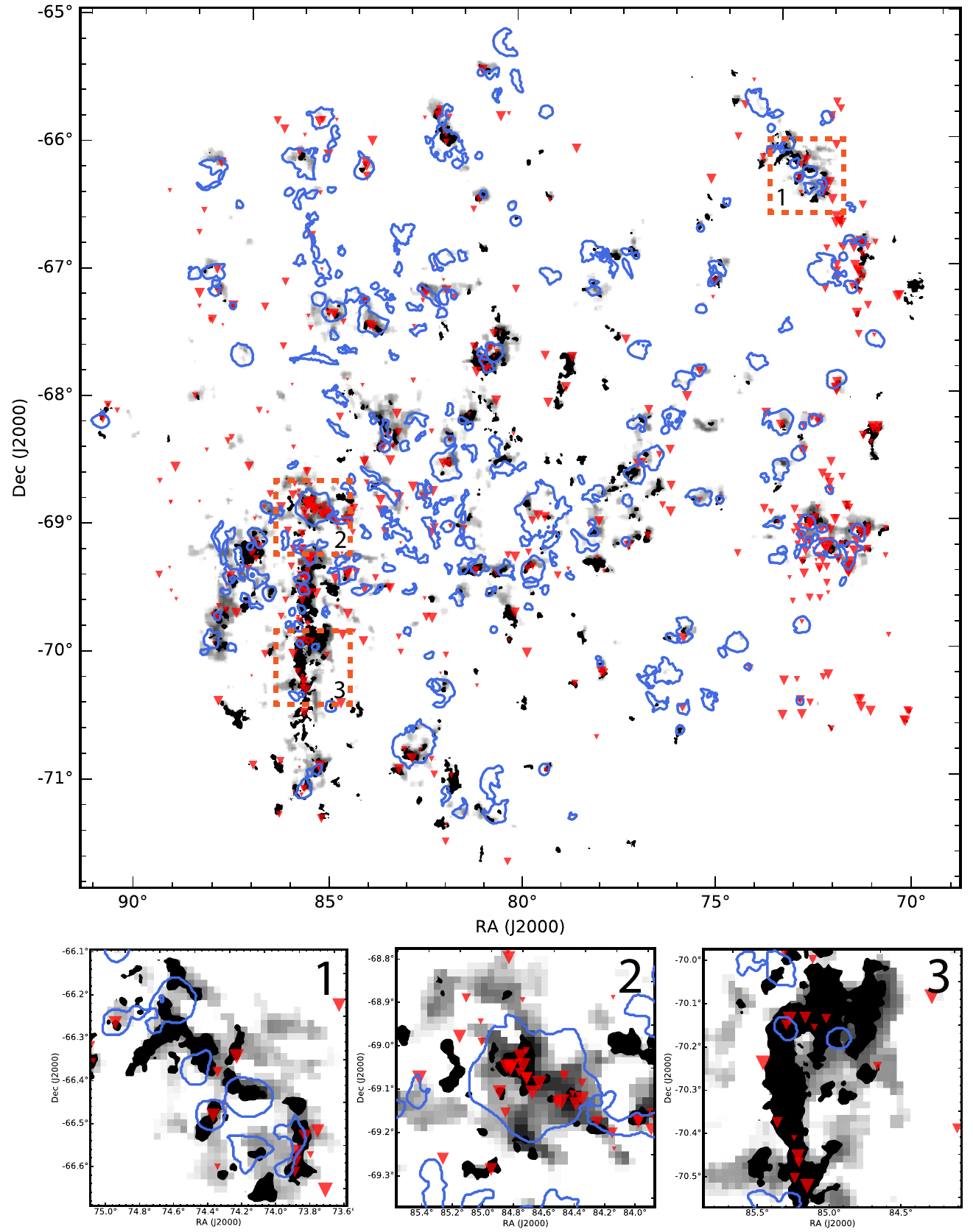} 
\caption{Overview of the LMC. Shown is the MAGMA ({\em filled black contours}; \citealt{wong_2011}) and the dust-based H$_\m{2}$ data {\em (grayscale; J16)}. Overplotted are the locations of Stage 1 MYSOs ({\em red inverted triangles}; \citealt{ochsendorf_2016}): the size of the symbols reflect total mass (Sec. \ref{sec:mysos}). Many MYSOs appear unassociated with GMCs; this likely originates from the limited coverage and sensitivity of the Mopra and {\em Herschel} surveys (Sec. \ref{sec:cloud}; \citealt{wong_2011, ochsendorf_2016}). The emission structures (tracing star forming regions) derived from the H$\alpha$ dendrogram are also shown ({\em blue contours}; Sec. \ref{sec:indirect}). The orange dashed squares show the outlines of the insets below: N11 {\em (inset 1)}, 30 Dor {\em (inset 2)}, and part of the South Molecular Ridge {\em (inset 3).}}
\label{fig:method}
\end{figure*}

\subsection{Cloud decomposition}\label{sec:cloud}
We use the Magellanic Mopra Assesment (MAGMA) DR3 (\citealt{wong_2011}, Wong et al., in prep) CO intensity map to determine molecular masses using $M_\m{mol}$ = $\alpha_\m{CO}$$L_\m{CO}$, where $L_\m{CO}$ is the CO luminosity and $\alpha_\m{CO}$ = 8.6 (K\,km\,s$^{-1}$\,pc$^2$)$^{-1}$ is the proportionality constant appropriate for the LMC \citep{bolatto_2013}. In addition, we use the dust-based molecular hydrogen map of \citet{jameson_2015} (from now: J16). The J16 map combines far-infrared dust emission (modeled with a single temperature blackbody modified by a broken power-law emissivity; \citealt{gordon_2014}) and atomic hydrogen maps to estimate the H$_\m{2}$ distribution. By doing so, J16 attempt to circumvent the biases of $^{12}$CO\,(1-0), which is known to probe a limited range in volume densities because of critical density, depletion, opacity, and photo-chemical effects. Furthermore, at the reduced metallicity of the LMC, a significant part of H$_\m{2}$ may be in a `CO-dark' phase \citep{madden_1997,leroy_2011}. We test the dependency of our results on the choice of molecular gas tracer (shown in Fig. \ref{fig:method}) by performing our analysis on both the MAGMA and J16 maps. 

We decompose the molecular cloud maps using the dendrogram technique \citep{rosolowsky_2008}. Dendrograms need three user-defined inputs in order to identify regions whilst minimizing contamination by spurious artifacts. First, a lower threshold below which data is to be excluded, which is set to the 2$\sigma$ sensitivity of our CO maps, $\sim$\,10 $M_\m{\odot}$ pc$^{-2}$ for MAGMA \citep{wong_2011}, and $\sim$\,15 $M_\m{\odot}$ pc$^{-2}$ for the dust-based H$_\m{2}$ map (J16). Second, a minimum surface density contrast for a structure to be considered as a separate entity, set to 1$\sigma$. Third, a minimum area each structure must subtend, set to twice the beam area of the MAGMA (45" FWHM) or J16 (1') maps. Uncertainties in molecular masses from MAGMA include the intrinsic noise of the data and the systematic uncertainty of the CO-to-H$_2$ conversion factor (accurate to within a factor of $\sim$\,2; \citealt{bolatto_2013}). We estimate the uncertainty introduced by the intrinsic noise of the data by adding random normally-distributed noise with amplitude 1$\sigma$ to the data: we find that the systematic uncertainty in the CO-to-H$_2$ conversion factor always dominates. Thus, the total uncertainty on molecular masses is estimated at $\pm$0.3 dex; a similar uncertainty was derived by J16 for the dust-based H$_\m{2}$ map. 

The total mass contained in the MAGMA clouds is 2.4\,$\times$\,10$^7$ M$_\m{\odot}$, which is a factor of two higher compared to the value of \citet{wong_2011}: this is because we use a higher $\alpha_\m{CO}$ factor of 8.6 (K\,km\,s$^{-1}$\,pc$^2$)$^{-1}$ versus 4.3 (K\,km\,s$^{-1}$\,pc$^2$)$^{-1}$ used by \citet{wong_2011}. The total mass found in the J16 map is 4.0\,$\times$\,10$^7$ M$_\m{\odot}$, consistent with values reported by \citet{jameson_2015}.

We provide cloud properties such as median mass ($\tilde{M}_\m{cloud}$), radius ($\tilde{R}_\m{cloud}$), mean density ($\tilde{\langle \rho \rangle}$), and free-fall time ($\tilde{\tau_\m{ff}}$) in Tab. \ref{tab:sfr}. The GMC free-fall time equals $\tau_\m{ff}$ = $\sqrt{3\pi/32G\rho}$, and $\rho$ = $M_\m{cloud}$/(4/3$\pi$$R^3$) its mean density. Between cloud Types, Type 3 clouds are larger and more massive compared to Type 1 and Type 2 clouds \citep{kawamura_2009}, while mean densities and free-fall times do not differ significantly. Comparing MAGMA and J16 clouds, one can see that J16 clouds are on average larger, more massive, and have lower internal densities (and higher free-fall times). This can also be discerned by examining Fig. \ref{fig:method}: despite the lower sensitivity of the J16 versus MAGMA map (see above), the J16 clouds appear more extended, suggesting that the GMCs in the LMC are surrounded by extensive diffuse `CO-dark' envelopes \citep{leroy_2011,bolatto_2013} that constitute roughly $\sim$\,50\% of molecular mass in the LMC.

\subsection{Star counting: MYSOs}\label{sec:mysos}

The most direct measure of SFR can be obtained through counting of Young Stellar Objects (YSOs), given by SFR = $N$(YSOs)\,$\times$\,$\langle{M}_\m{\star}\rangle$ / $t_\m{\star}$, where the average mass is $\langle{M}_\m{\star}\rangle$ $\approx$ 0.5 for a fully-sampled initial mass function \citep{kroupa_2001}, and $t_\m{\star}$ is the age of the YSO population. This technique has been applied to the nearest ensemble of molecular clouds \citep{evans_2009,heiderman_2010,lada_2010}. However, counting low-mass YSOs becomes cumbersome at large distances because of completeness and/or crowding. This can be overcome by considering only the most luminous sources: massive YSOs (MYSOs). \citet{ochsendorf_2016} present a MYSO catalog that builds on the results of galaxy-wide searches of MYSOs in the LMC \citep{whitney_2008,gruendl_2009,seale_2014}. This catalogue has been estimated to be complete for Stage 1 MYSOs of mass $M$\,$\textgreater$\,8 $M_\m{\odot}$ (Fig. \ref{fig:method}). We utilize this catalogue to obtain a census of massive star formation by counting the number of MYSOs in each cloud identified in the dendrogram decomposition (Sec. \ref{sec:cloud}). We assume that the observed luminosity is dominated by a single massive source (see below), and use the \citet{robitaille_2006} models to obtain a source mass (see \citealt{robitaille_2008} for a complete discussion of this conversion). Subsequently, we multiply the obtained source mass with an IMF \citep{kroupa_2001} to account for stars below our completeness limit. For $t_\m{\star}$, we choose 0.5 Myr, which is the most recent value obtained for the observationally-derived `Class 1' low-mass sources (which largely overlap with the theoretically-based `Stage 1' sources; \citealt{heiderman_2015, heyer_2016}) in the Gould's Belt \citep{dunham_2015}. It is not clear whether this value applies to massive stars; however, it should represent a reasonable first-order estimate of $t_\m{\star}$. Our completeness limit of $M$\,$\textgreater$\,8 $M_\m{\odot}$ for MYSOs then translates to a lower limit of SFR$_\m{MYSO}$\,$\sim$\,100 $M_\m{\odot}$ Myr$^{-1}$.

In many cases, our MYSO sources will break into small clusters at higher resolution \citep{vaidya_2009,stephens_2017}. To estimate the uncertainty in our SFR$_\m{MYSO}$ measurement, we consider the case if we were to observe the Orion Trapezium cluster at the distance of the LMC, where the reprocessed IR luminosity would appear as a compact source in our IR maps. Using the stellar atmosphere models of \citet{vacca_1996} we estimate that the main ionizing source, $\theta^1$ Ori C, contains $\simeq$\,50\% of the total luminosity of the Trapezium \citep{o'dell_1993,simon-diaz_2003}. Thus, our measured IR luminosity would overestimate the luminosity of the most massive source by $\sim$\,0.3 dex, which translates to an overestimation of the stellar mass by a factor of $\sim$\,0.1 dex \citep{mottram_2011}; we adopt this as our systematic uncertainty in SFR$_\m{MYSO}$.

\subsection{Indirect SFR tracers: H$\alpha$ and 24 $\mu$m}\label{sec:indirect}
Counting Stage 1 MYSOs (Sec. \ref{sec:mysos}) provides a census of the young, embedded phase of massive star formation. In contrast, the widely-used SFR diagnostic H$\alpha$ traces a more evolved population of massive stars of 3 - 10 Myr \citep{kennicutt_2012}. Given its sensitivity to dust attenuation, we correct the H$\alpha$ emission (from the Southern H-Alpha Sky Survey Atlas, SHASSA; \citealt{gaustadt_2001}) for extinction using 24 $\mu$m emission (from Spitzer's Surveying the Agents of a Galaxy's Evolution, SAGE; \citealt{meixner_2013}). Emission at 24 $\mu$m traces star formation activity up to 100 Myr, but probably less on small scales if associated with H$\alpha$. In this case, young massive stars will dominate ionization and local dust heating (for a discussion, see \citealt{vutisalchavakul_2013}). 

We convolve the 24 $\mu$m map to the resolution of our H$\alpha$ map (0.8'), and use the \citet{calzetti_2007} relationship to transform H$\alpha$ and 24 $\mu$m luminosity, $L$\,(H$\alpha$) and $L$\,(24$\mu$m), to a SFR:

\begin{equation}	
\begin{split}
\m{SFR}_\m{H\alpha}\,(M_\m{\odot}\,\m{yr}^{-1}) = 5.3\times10^{-42} [L\,(\m{H}\alpha) \\ 
+ \,0.031\,L\,(24\,\mu\m{m})].
\end{split}
\label{eq:sfr}
\end{equation}

\citet{koepferl_2016} have shown that the 24 $\mu$m SFR tracer alone can significantly underestimate the SFR. However in the LMC, 24 $\mu$m typically contributes only $\lesssim$\,20\% to the measured SFR (see also J16). We again use dendrograms \citep{rosolowsky_2008} to characterize the diffuse emission across the face of the LMC and automate the identification of star formation regions (Fig. \ref{fig:method}). The H$\alpha$ observations show that the diffuse background in the LMC has values ranging between 10 - 200 R (Rayleigh; 1 R = 10$^6$/4$\pi$ photons s$^{-1}$ cm$^{-2}$ sr$^{-1}$), with a typical value of $\sim$\,100 R. We thus use 300 R as a lower threshold for our dendrogram, 150 R as a minimum surface brightness contrast, and an minimum area of 5 times the beam size of SHASSA ($\sim$\,48"). We choose this rather large minimum area since the SHASSA map shows artifacts of imperfect subtraction of bright (foreground) stars, which are typically several pixels in size. These parameters set the detection limit of SFR$_\m{H\alpha}$: using Eq. \ref{eq:sfr}, a star formation region of 5 SHASSA beam areas filled with a surface brightness of 300 R (i.e., the adopted lower threshold) contains a SFR$_\m{H\alpha}$\,$\sim$\,5 $M_\m{\odot}$ Myr$^{-1}$. Note that only stars with spectral type B2 or earlier can maintain an \HII\ region, which intrinsically sets a lower limit to SFR$_\m{H\alpha}$. With an ionizing flux of $\sim$\,10$^{47.5}$ photons s$^{-1}$ \citep{schaerer_1997} and assuming all ionizing photons get absorbed, we estimate the expected H$\alpha$ luminosity (assuming case B recombination; \citealt{osterbrock_2006}), and thereby SFR$_\m{H\alpha}$ through Eq. \ref{eq:sfr}. We find that an \HII\ region powered by a single B2V star should exhibit SFR$_\m{H\alpha}$\,$\sim$\,2.5 $M_\m{\odot}$ Myr$^{-1}$, comparable to our sensitivity limit. 
 
By summing the H$\alpha$ luminosities of all star forming regions we recover $\sim$\,50\% of the total H$\alpha$ luminosity of the entire LMC. This is consistent with an escape fraction of ionizing photon luminosity from \HII\ regions of $\sim$\,0.5 observed by \citet{pellegrini_2012}. Consequently, we correct for the effects of ionizing photon leakage by multiplying our SFR$_\m{H\alpha}$ measurements by a factor of 2.

We note that Eq. \ref{eq:sfr} assumes a fully-sampled IMF, and averaging over large enough spatial scales so that each phase of the star forming process, i.e., from the deeply embedded phase to fully exposed clusters, is adequately probed \citep{krumholz_2014}. At small scales \citep{kruijssen_2014} or at low SFR \citep{dasilva_2012}, these conditions are likely to be violated, which will introduce scatter and a potential bias in the derived SFR$_\m{H\alpha}$ values because of stochastic sampling of the high-end tail of the IMF. These effects become important below cluster masses of $\sim$\,3000 M$_\odot$ \citep{cervino_2003}. To estimate the stellar masses associated with our observed SFR$_\m{H\alpha}$, we relate the H$\alpha$ luminosity to ionizing photon rate $Q_\m{0}$ through $L$(H$\alpha$) = $\frac{\alpha_\m{H\alpha}^{\m{eff}}}{\alpha_\m{B}}$$E_\m{H\alpha}$ $\approx$ 1.37$\times$10$^{-12}$$Q_\m{0}$, where $\alpha_\m{H\alpha}^{\m{eff}}$ is the effective recombination coefficient at H$\alpha$, $\alpha_\m{B}$ is the case B recombination coefficient, and $E_\m{H\alpha}$ the energy of an H$\alpha$ photon \citep{osterbrock_2006}. From this, we estimate the stellar mass through $M_\m{\star}$ = 6.3$\times$10$^{-46}$$Q_\m{0}$, where the constant on the right hand represents the IMF averaged photon rate divided by the IMF averaged stellar mass (in units of s$^{-1}$\,M$_\odot$; \citealt{murray_2011}). It follows that the median stellar mass of our H$\alpha$ emitting star forming regions is  $\sim$\,10$^3$ M$_\m{\odot}$, implying that a significant portion of our SFR measurements will be affected by stochasticity.

We use the `Stochastically Lighting Up Galaxies' tool (SLUG; \citealt{krumholz_2014}) to estimate the bias and scatter (defined in \citealt{dasilva_2012}) associated with stochasticity for our SFR$_\m{H\alpha}$ measurements. We use the default SLUG library \citep{dasilva_2012} at solar metallicity and a 500 Myr continuous star formation scale. Note that the metallicity of the LMC is roughly half solar \citep{russell_1992}. In this regard, metallicity induces variations in the conversion between H$\alpha$ luminosity and SFR on the order of $\pm$0.1 dex \citep{calzetti_2007}. In addition, on the scale of individual \HII\ regions, a $\textless$\,10 Myr star formation timescale may be more appropriate. However, assuming a constant star formation rate, the H$\alpha$ luminosity per unit star formation rate reaches a steady state after only a few Myr (as stars arriving on the zero age main sequence are balanced by ones that evolve toward supernovae; \citealt{krumholz_2007, kennicutt_2012}). Nonetheless, on the scale of individual star forming regions, the star formation rate is likely not continuous, but will proceed in discreet bursts \citep{faesi_2014}. In this regard, the ubiquity of `two-stage' starbursts \citep{walborn_1992} in the LMC \citep{ochsendorf_2016} shows that massive star formation typically clusters over multiple generations on timescales up to at least $\sim$\,10 Myr. Depending on the amount of subclusters formed, approximating the star formation history as continuous may be a reasonable first order estimate. Finally, uncertainties in our flux measurements are estimated at $\pm$0.05 dex. 

The total scatter in our SFR$_\m{H\alpha}$ measurements can be estimated by adding in quadrature the uncertainties from stochasticity (evaluated at log\,[SFR$_\m{H\alpha}$ (M$_\odot$ Myr$^{-1}$)] = 2.8, the median of Type 2 and Type 3 clouds), metallicity, and photometry. We do not consider the bias induced by stochasticity, which is generally small at log\,[SFR$_\m{H\alpha}$ (M$_\odot$ Myr$^{-1}$)] = 2.8 compared to the estimated total scatter of our measurements, $\pm$\,0.4 dex (a similar scatter is found by comparing the SFR$_\m{MYSO}$ and SFR$_\m{H\alpha}$ tracers; see Fig. \ref{fig:sfr_tracers} and Sec. \ref{sec:results}).

\subsection{Matching GMCs with SFR tracers}\label{sec:match}

We aim to derive the SFR associated with each individual GMC found in our cloud decompositions (Sec. \ref{sec:cloud}). For the MYSOs, we simply add all MYSOs found within the footprint of a GMC, from which we calculate SFR$_\m{MYSO}$ (Sec. \ref{sec:mysos}). For H$\alpha$, we cross-match each structure defined by our H$\alpha$ dendrogram (Sec. \ref{sec:indirect}) with our GMC dendrogram, and define a match when there is physical overlap between both structures. We then calculate SFR$_\m{H\alpha}$, taking into account the local 24 $\mu$m emission (Sec. \ref{sec:indirect}). If there is more than one emission structure per GMC, we sum the single components. Likewise, if there are more GMCs found within a given emission structure, we add the single cloud components. We perform the above described routine for the MAGMA and J16 molecular maps, and ultimately obtain a list of GMCs (with $M_\m{cloud}$) with associated SFR$_\m{MYSO}$ and SFR$_\m{H\alpha}$. 

\subsection{Cloud classification}\label{sec:classification}

\citet{kawamura_2009} classified GMCs in the LMC as Type 1 (GMCs with no massive star formation), Type 2 (GMCs with associated \HII\ regions), and Type 3 (GMCs with associated \HII\ regions and optical stellar clusters). This classification was based on GMCs detected in the NANTEN survey (at resolution 2.6'; \citealt{fukui_2008}). As in \citet{ochsendorf_2016}, we match our GMC lists with the \citet{kawamura_2009} catalogue. For MAGMA, we consider all GMCs detected within the footprint of a NANTEN GMC as being of the same `Type'. The J16 clouds are typically more extended than the NANTEN clouds of \citet{kawamura_2009} and, in some cases, individual NANTEN clouds of different Types are joined in a single structure. In these cases, we label the cloud as the highest `Type' (i.e., most evolved) found within the individual NANTEN components. 

\subsection{A word on star formation efficiencies \& final products}\label{sec:classification}
Star formation efficiencies are often defined differently in extragalactic and Galactic studies, which stems from the (in)capability to resolve individual stars, clusters, and clouds in specific environments. Given that the LMC represents a special case in this sense (i.e., both extragalactic and resolved in individual star forming regions), we consider both cases, as this could be useful for intercomparisons across extragalactic and Galactic studies. We follow the notation of \citet{kennicutt_2012}, and write $\epsilon'$ = SFR/$M_\m{cloud}$, which is often used in extragalactic studies. In addition, we are particularly interested in the star formation efficiency per free-fall time, $\epsilon_\m{ff}$ = $\tau_\m{ff}$/$t_\m{dep}$ = $\tau_\m{ff}$\,$\times$\,$\epsilon'$ (where $t_\m{dep}$ is the depletion timescale). 

We provide the results of our cloud decomposition and SFR measurements in four machine-readable tables in the appendix: (1) MAGMA clouds and MYSOs, (2) MAGMA clouds and Ha\,+\,24\,$\mu$m, (3) J16 clouds and MYSOs, and (4) J16 clouds and Ha\,+\,24\,$\mu$m.

\section{Results}\label{sec:results}

\subsection{Star formation rate tracers}

\begin{figure}
\centering
\includegraphics[width=8.5cm]{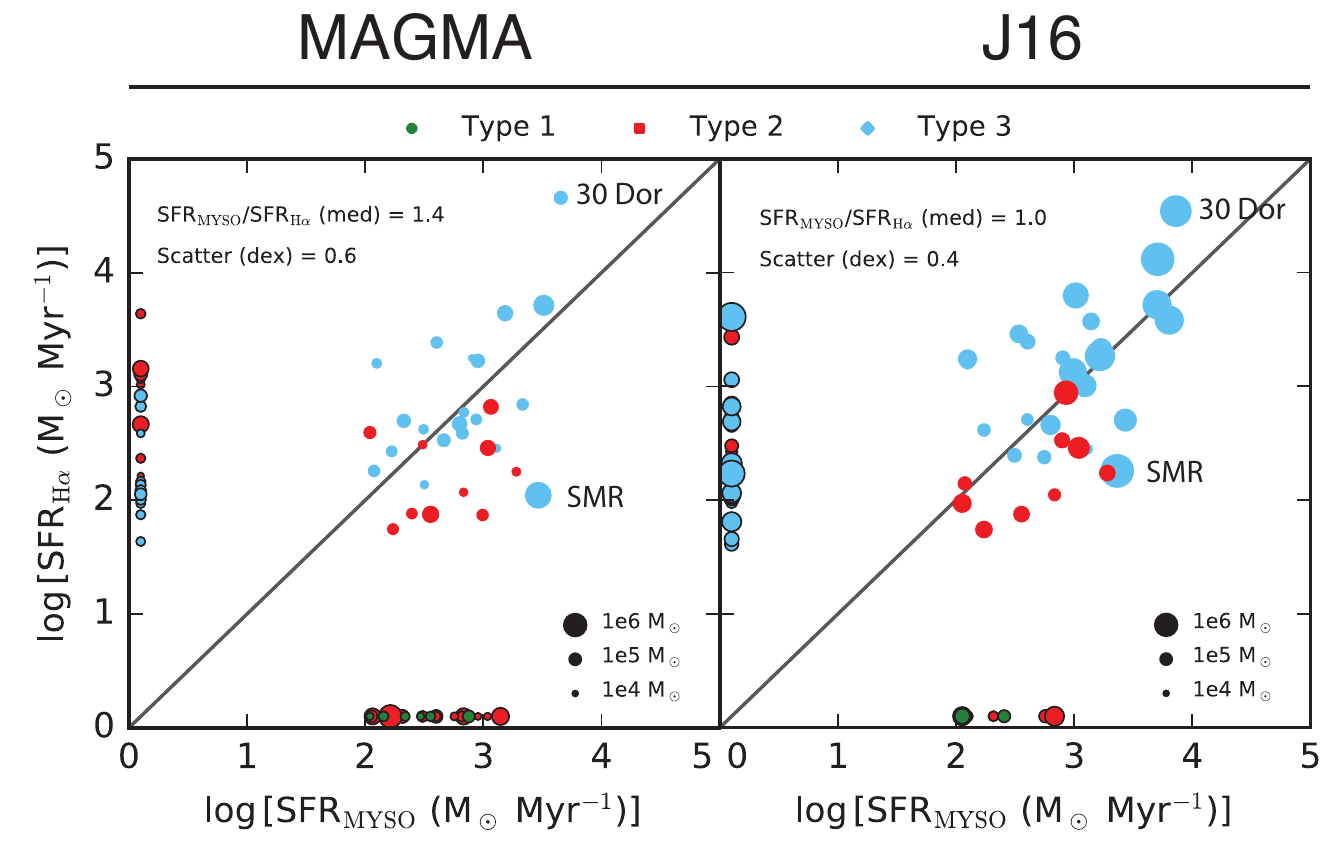} 
\caption{{\bf (a)}: Comparison of SFR tracers. Plotted is SFR$_\m{MYSO}$ versus SFR$_\m{H\alpha}$  in the MAGMA {\em (left panel)} and J16 {\em (right panel)} decomposition for Type 1 ({\em green}), Type 2 ({\em red}), and Type 3 ({\em light blue}) clouds. The solid line marks where both tracers are equal. Shown is the median value of SFR$_\m{MYSO}$/SFR$_\m{H\alpha}$ and its scatter in dex. The size of the symbols reflect the GMC mass of each star forming region.}
\label{fig:sfr_tracers}
\end{figure}

Fig. \ref{fig:sfr_tracers} compares SFR$_\m{MYSO}$, SFR$_\m{H\alpha}$, and $M_\m{cloud}$; a similar comparison has earlier been made for small samples in the Milky Way \citep{chomiuk_2011,lee_2016}. Here, we extend these studies by sampling massive star forming regions on a galaxy-wide scale. Assuming $t_\star$ = 0.5 Myr (Sec. \ref{sec:mysos}), we find good agreement between SFR$_\m{MYSO}$ and SFR$_\m{H\alpha}$, with a mean ratio of 1.4 and 1.0 for the MAGMA and J16 decomposition, respectively. Nonetheless, there is considerable scatter between both SFR tracers, i.e., $\sim$\,0.6 and $\sim$\,0.4 dex for the MAGMA and J16 clouds. This is only slightly larger from what is expected from stochastic sampling of the IMF alone (Sec. \ref{sec:indirect}), although the exact extent of this scatter is known to vary with SFR \citep{dasilva_2012,krumholz_2014}. If stochasticity dominates the scatter in Fig. \ref{fig:sfr_tracers}, this would in its turn imply that SFR$_\m{MYSO}$ is a robust, unbiased tracer of SFR on individual cloud scales: the largest uncertainty remains $t_\star$ \citep{kennicutt_2012}. 

A substantial fraction of clouds are only observed in a single SFR tracer (Fig. \ref{fig:sfr_tracers}), which may be expected given that we are sampling star formation regions in different evolutionary states \citep{kawamura_2009,kruijssen_2014}. Star formation regions captured in the earliest stages of evolution may only be detected through (massive) YSOs, whilst regions with fully exposed clusters may no longer have MYSOs as the molecular gas reservoir is disrupted. In this regard, simulations show that stellar feedback quickly clears cavities around young star clusters ($\gtrsim$\,1 Myr) depending on the initial properties of the cloud \citep{dale_2012,dale_2014}. On the other hand, observations may imply longer timescales ($\gtrsim$\,5\,-\,10 Myr) for clusters to clear their immediate surroundings of CO emitting gas \citep{leisawitz_1989,kawamura_2009}. The sensitivity of surveys targeting the molecular gas also plays a role here, which may cause MYSOs to fall outside CO cloud footprints \citep{wong_2011}. Consistent with the above, we find a ratio of SFR$_\m{MYSO}$/SFR$_\m{H\alpha}$\,$\sim$\,3 for Type 2 clouds (both MAGMA and J16 clouds), while this ratio drops below unity for Type 3 clouds. However, the mean ratio of SFR$_\m{MYSO}$/SFR$_\m{H\alpha}$ over the entire LMC shows that, averaged over the lifetime of star forming regions, both SFR tracers are consistent with one another.

\subsection{Massive star formation rates and efficiencies}\label{sec:rates}

\begin{figure*}
\centering
\includegraphics[width=18cm]{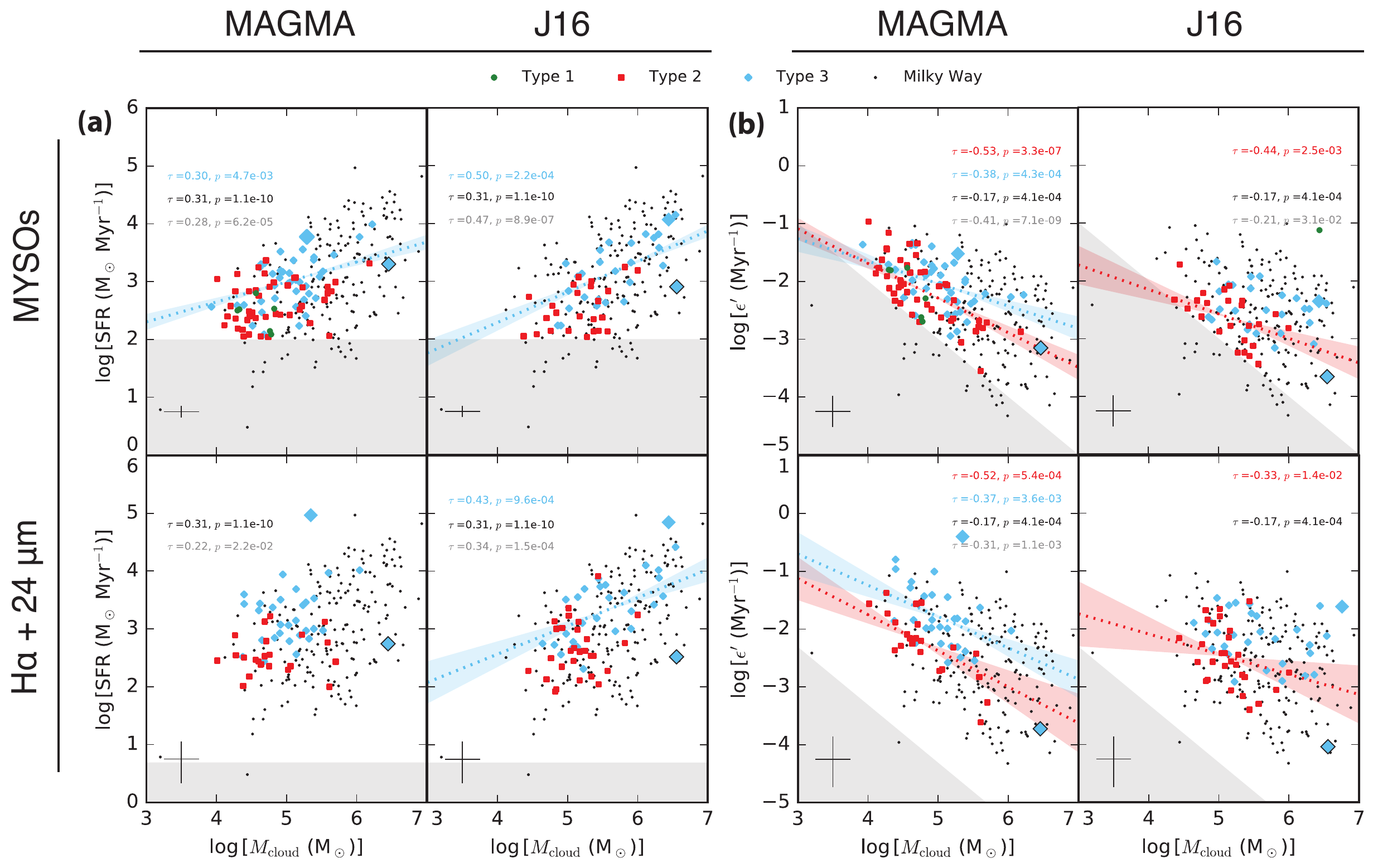} 
\caption{{\bf (a)}: Star formation rate (SFR) versus cloud mass ($M_\m{cloud}$). Results are shown for SFR$_\m{H\alpha}$ and SFR$_\m{MYSO}$ for both MAGMA and J16 clouds. Plotted are Type 1 clouds ({\em green circles}), Type 2 clouds ({\em red squares}), Type 3 clouds ({\em blue diamonds}), and the Galactic sample from \citet{lee_2016} ({\em black dots}). Type 1 clouds are only detected in SFR$_\m{MYSO}$ \citep{ochsendorf_2016}, but not in SFR$_\m{H\alpha}$ \citep{kawamura_2009}. The J16 cloud map does not have the sensitivity to detect Type 1 clouds, which generally have low surface densities. Gray shaded areas represents regions below the lower limit to which we can observe the SFR tracer (i.e., low-mass star formation; Secs. \ref{sec:mysos} \& \ref{sec:indirect}). The position of 30 Doradus ({\em oversized diamond}) and the South Molecular Ridge ({\em oversized diamond with black edge}) are shown. The estimated systematic uncertainty (defined as the range between the 16th and 84th percentile of the associated probability density functions) on both axes are shown in lower left, derived by considering various sources of scatter (Secs. \ref{sec:cloud}, \ref{sec:mysos}, and \ref{sec:indirect}). 
Kendall rank correlation coefficients are shown for cloud subsets that have a statistically significant correlation (i.e., $p$-value\,$\textless$\,0.05). {\bf (b):} Same, but for the star formation efficiency $\epsilon'$ = SFR/$M_\m{cloud}$.}
\label{fig:sfe}
\end{figure*}

\begin{figure}
\centering
\includegraphics[width=8.5cm]{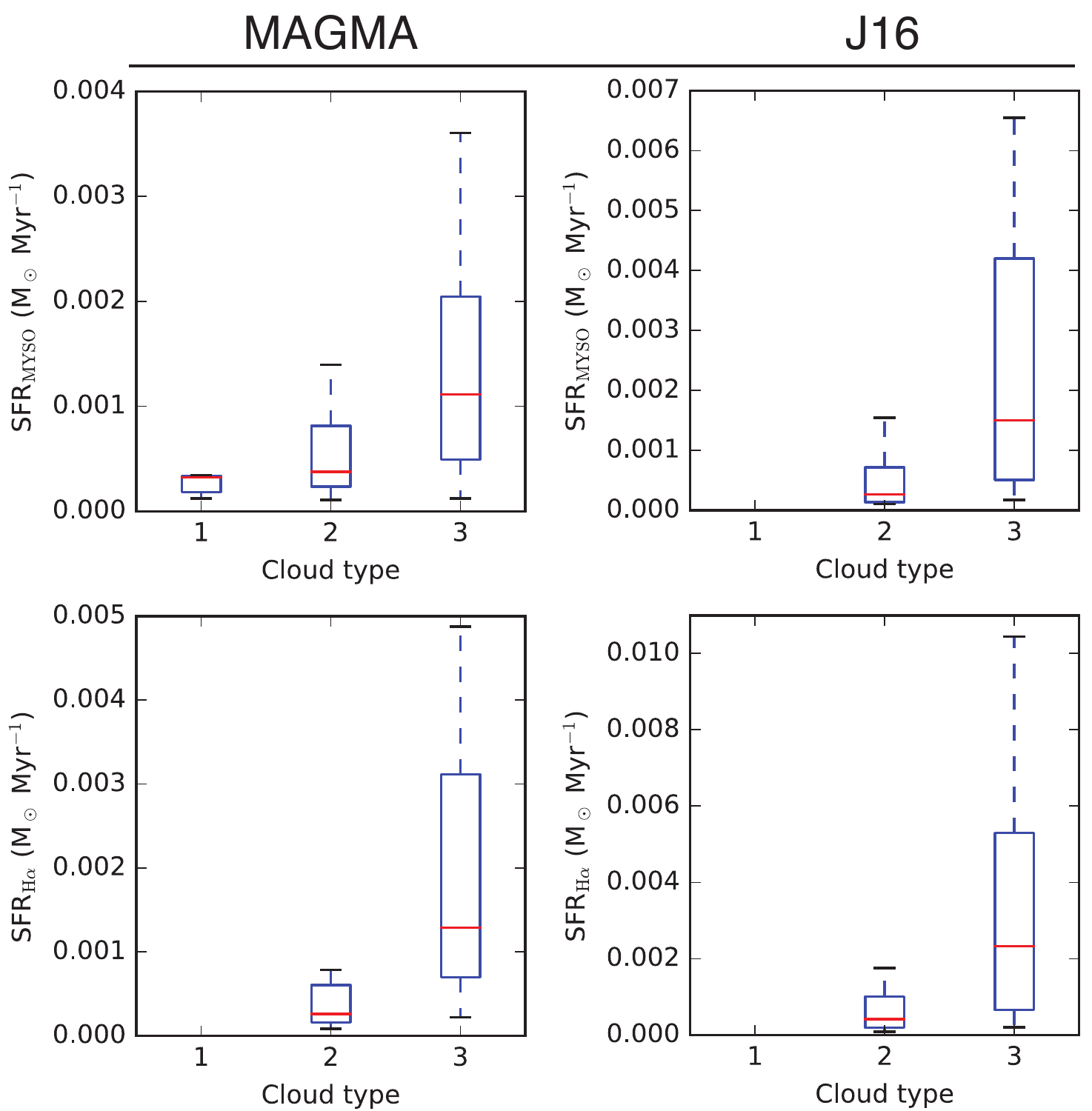} 
\caption{Whisker plots of SFR$_\m{MYSO}$ and SFR$_\m{H\alpha}$ for MAGMA (left) and J16 (right). Note the change of scale between individual plots. Absolute rates increase with cloud Type in both star formation indicators.}
\label{fig:boxplot}
\end{figure}

Fig. \ref{fig:sfe}a plots SFR versus $M_\m{cloud}$ for both tracers (SFR$_\m{MYSO}$ and SFR$_\m{H\alpha}$) and both cloud decompositions (MAGMA and J16). We recover systematically higher SFR$_\m{H\alpha}$ {\em and} SFR$_\m{MYSO}$ along the Type 1 - Type 3 evolutionary sequence \citep{kawamura_2009}. We emphasize this result in a whisker plot (Fig. \ref{fig:boxplot}), and provide absolute numbers in Tab. \ref{tab:sfr}. In addition, the fraction of clouds containing MYSOs, $p_\m{MYSO}$, increases steadily between Type 1, Type 2, and Type 3 clouds. Thus, it becomes clear that in the presence of stellar clusters (i.e., Type 2 and Type 3 clouds) SFR$_\m{H\alpha}$ increases, which appears to go hand-in-hand with an increased star formation rate of younger generations of massive stars as reflected by SFR$_\m{MYSO}$ \citep[see also][]{ochsendorf_2016}. 

We fit a linear model log[SFR] = log[$M_\m{cloud}$]$\alpha$ + $\beta$ to the datapoints of Fig. \ref{fig:sfe}a. We perform a linear least-squares fit on three different subsets: Type 2, Type 3, and the entire set of clouds. To estimate the uncertainties on our fit, we approximate the errors on $M_\m{cloud}$, SFR$_\m{MYSO}$, and SFR$_\m{H\alpha}$ as being normally-distributed with amplitude 0.3, 0.1, and 0.4 dex, respectively (Secs. \ref{sec:cloud},  \ref{sec:mysos}, and \ref{sec:indirect}). We subsequently add random noise to our data, and perform the fit 100 times. Table \ref{tab:sfr} present the mean and 1$\sigma$ uncertainty on the derived slope $\alpha$ in case a statistically significant correlation is found (i.e., Kendall rank correlation $p$-value $\textless$\,0.05; Fig. \ref{fig:sfe}). In all cases, we find that SFR$_\m{MYSO}$ and SFR$_\m{H\alpha}$ depend on $M_\m{cloud}$  {\em sub-linearly} for both the MAGMA and J16 clouds. 

The 30 Doradus region contains the largest and most massive \HII\ region in the Local Group (harboring $\sim$\,2400 OB stars; \citealt{parker_1993}) and is often considered the nearest super star cluster that is reminiscent of starburst environments \citep{walborn_1991}. Conversely, the South Molecular Ridge (SMR) is a GMC which contains about 30\% of the total molecular mass of the LMC, yet exhibits a remarkably low star formation rate \citep{indebetouw_2008}. Given the extremities in GMC and star formation behavior of both these systems, we illustrate how our results are affected by these systems. Table \ref{tab:sfr} lists fit results for $\alpha_\m{SFR,H\alpha}$ and $\alpha_\m{\epsilon',H\alpha}$ (see below) by excluding 30 Doradus and the SMR. We note that $\alpha_\m{SFR,MYSO}$ and $\alpha_\m{\epsilon',MYSO}$ are not significantly affected in this way, as 30 Doradus and the SMR do not appear as outliers in the SFR$_\m{MYSO}$ tracer.

We compare our results with the Galactic sample of \citet{lee_2016} in Fig. \ref{fig:sfe}a and Tab. \ref{tab:sfr}. We find good agreement between both galaxies. For example, the star forming regions in the LMC and Milky Way occupy the same region in `$M_\m{cloud}$\,-\,SFR' space. The Milky Way does have a larger dynamic range on both axes, which may explain some of the offset in median SFR between the MAGMA clouds and the Milky Way (a factor of $\sim$\,1.5; Tab. \ref{tab:sfr}). However, the derived slopes of $\alpha_\m{SFR,MYSO}$ and $\alpha_\m{SFR,H\alpha}$ all agree within $\sim$\,1$\sigma$ for the J16 clouds and the Milky Way, showing that a sublinear relation between SFR and {\em total} cloud mass persist in both the LMC and the Milky Way \citep{vutisalchavakul_2016,lee_2016}. 

\begin{table*}
\centering
\caption{Cloud and star formation properties}
\begin{tabular}{l||c|c|c|c||c|c|c|c||c}\hline

& \multicolumn{4}{c||}{\bf MAGMA} & \multicolumn{4}{c||}{\bf J16} & {\bf MW}  \\ \hline
& Type 1 & Type 2 & Type 3 & All  & Type 1 & Type 2 & Type 3 & All & \\  \hline
$\tilde{M}_\m{cloud}$ (10$^5$ $M_\odot$) & 0.27 & 0.33 & 0.84 & 0.40 & 0.73 & 1.04 & 1.74 & 1.28 & 4.8 \\
$\tilde{R}_\m{cloud}$ (pc) & 19.9 & 20.6 & 28.3 & 21.5 & 34.2 & 36.7 & 44.6 & 39.0 & 43 \\ 
$\tilde{\langle \rho \rangle}$ (cm$^{-3}$) & 35.2 & 38.2 & 30.1 & 36.4 & 19.7 & 21.1 & 13.8 & 20.1 & 59.5 \\
$\tilde{\tau_\m{ff}}$ (Myr) & 8.7 & 8.3 & 8.8 & 8.5 & 11.6 & 11.2 & 12.2 & 11.5 & 6.7 \\
$\tilde{v}_\m{CO}$ (km s$^{-1}$) & 1.4 & 1.3 & 1.9 & 1.5 & & & & &  4.8 \\ 
$p_\m{MYSO}$ & 0.10 & 0.43 & 0.66 & 0.42 & & 0.47 & 0.69 & 0.52 \\ \hline

$\tilde{\m{SFR}}_\m{MYSO}$ (10$^{3}$ $M_\odot$/Myr) & 0.3 & 0.4 & 1.1 & 0.7 &  & 0.3 & 1.5 & 0.7 & \\
$\tilde{\m{SFR}}_\m{H\alpha}$\,\,\,\,\,\,\,\, (10$^{3}$ $M_\odot$/Myr) &  & 0.3 & 1.3 & 0.8 &  & 0.4 & 2.3 & 0.9 & 1.1$^{(\ast\ast)}$  \\ 
$\alpha_\m{SFR,MYSO}$ & & & 0.35$\pm$0.07 & 0.30$\pm$0.04 &  & 0.51$\pm$0.08 & 0.57$\pm$0.06 & & \\
$\alpha_\m{SFR,H\alpha}$ &  &  &  & 0.26$\pm$0.13 &  &  & 0.47$\pm$0.15 & 0.55$\pm$0.12 & 0.65 \\
$\alpha_\m{SFR,H\alpha}$$^{(*)}$  &  &  &  & 0.28$\pm$0.15 &  &  & 0.53$\pm$0.17 & 0.56$\pm$0.15 & 0.65 \\

$\alpha_\m{\epsilon',MYSO}$ &  &  -0.60$\pm$0.10 & -0.39$\pm$0.12  & -0.43$\pm$0.06 &  & -0.41$\pm$0.16 & & -0.21$\pm$0.09 &  \\ 
$\alpha_\m{\epsilon',H\alpha}$ &  &  -0.61$\pm$0.22 & -0.61$\pm$0.18  & -0.50$\pm$0.13 &  & -0.42$\pm$0.29 & &  & -0.35 \\ 
$\alpha_\m{\epsilon',H\alpha}$$^{(*)}$  &  &  -0.59$\pm$0.27 & -0.44$\pm$0.21  & -0.38$\pm$0.15 &  & -0.37$\pm$0.25 & &  & -0.35 \\ \hline

\end{tabular}
\tablecomments{Cloud and star formation properties for the LMC clouds (both MAGMA and J16) and the Milky Way (MW). Listed are: the median cloud mass ($\tilde{M}_\m{cloud}$); cloud radius ($\tilde{R}_\m{cloud}$); cloud density ($\tilde{\langle \rho \rangle}$); cloud free-fall time ($\langle\tau_\m{ff}\rangle$); velocity dispersion ($\tilde{v}_\m{CO}$); the fraction of clouds that contain MYSOs, $p_\m{MYSO}$; the median SFR$_\m{MYSO}$ ($\tilde{\m{SFR}}_\m{MYSO}$); the median SFR$_\m{H\alpha}$ ($\tilde{\m{SFR}}_\m{H\alpha}$); and the various slope indices $\alpha$ between SFR and $\epsilon'$ with $M_\m{cloud}$. $^{(*)}$: These values refer to the fits by excluding the 30 Doradus and SMR region (see text). $^{(\ast\ast)}$: this number is derived from free-free flux as opposed to H$\alpha$ + 24\,$\mu$m.}
\label{tab:sfr}
\end{table*} 
  
Figure \ref{fig:sfe}b plots the corresponding $\epsilon'$ of individual star forming complexes. Naturally, a higher SFR$_\m{MYSO}$ and SFR$_\m{H\alpha}$ leads to increased values of $\epsilon'_\m{MYSO}$ and  $\epsilon'_\m{H\alpha}$ along the Type 1 - Type 3 evolutionary sequence. However, the most striking result from Fig. \ref{fig:sfe}b is the decline of $\epsilon'_\m{H\alpha}$ and {\bf  $\epsilon'_\m{MYSO}$} with $M_\m{cloud}$, which holds for the MAGMA and J16 clouds, and is readily recognized in the Galactic GMCs. Thus, larger GMCs have higher depletion times ($\tau_\m{dep}$ = 1/$\epsilon'$): we defer a further discussion of the implications of this observation to Sec. \ref{sec:discussion}.

\begin{figure}
\centering
\includegraphics[width=8.5cm]{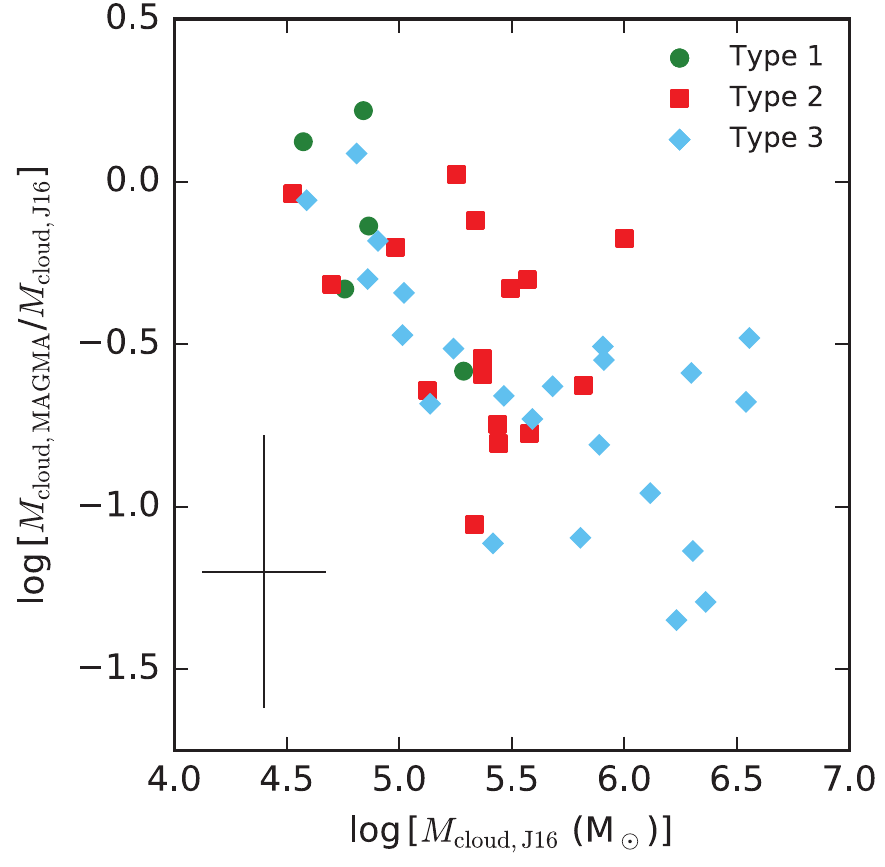} 
\caption{Comparison of mass found in the MAGMA and J16 clouds complexes. Only detections above 15 $M_\m{\odot}$ pc$^{-2}$ are taken into account to set the sensitivity of the MAGMA and J16 maps at the same level (Sec. \ref{sec:cloud}). The typical error is indicated in the lower left, estimated by adding the individual errors on mass in quadrature.}
\label{fig:gmc_comp}
\end{figure}

From Larson's laws \citep{larson_1981}, it is known that more massive clouds have lower $\tilde{\langle \rho \rangle}$, which may impact the star formation efficiency of GMCs. While we see a slight tendency for $\tilde{\langle \rho \rangle}$ to decrease along Type 1 - Type 3 clouds (Tab. \ref{tab:sfr}), this difference is not significant given the uncertainties on cloud mass ($\sim$\,0.3 dex; Sec. \ref{sec:cloud}. Alternatively, we may explore the structure of individual clouds by plotting the ratio $M_\m{cloud,MAGMA}/M_\m{cloud,J16}$, since the $M_\m{cloud,J16}$ values are more likely to include diffuse gas (Sec. \ref{sec:cloud}). Here, we only take into account mass above 15 $M_\m{\odot}$ pc$^{-2}$, i.e., the sensitivity limit for the dust-based H$_\m{2}$ map (J16). While the predicted errors are large, Fig. \ref{fig:gmc_comp} shows that this ratio declines as a function of $M_\m{cloud,J16}$. This implies that larger clouds have larger diffuse envelopes, and grow beyond $\sim$\,10$^5$ mostly by adding 'CO-dark' gas. The decline of $\epsilon'$ may therefore be caused by an increasing `contamination' of GMCs by addition of diffuse gas at higher $M_\m{cloud}$.

\subsection{The star formation efficiency per free-fall time}\label{sec:eff}
 
\begin{figure*}
\centering
\includegraphics[width=18cm]{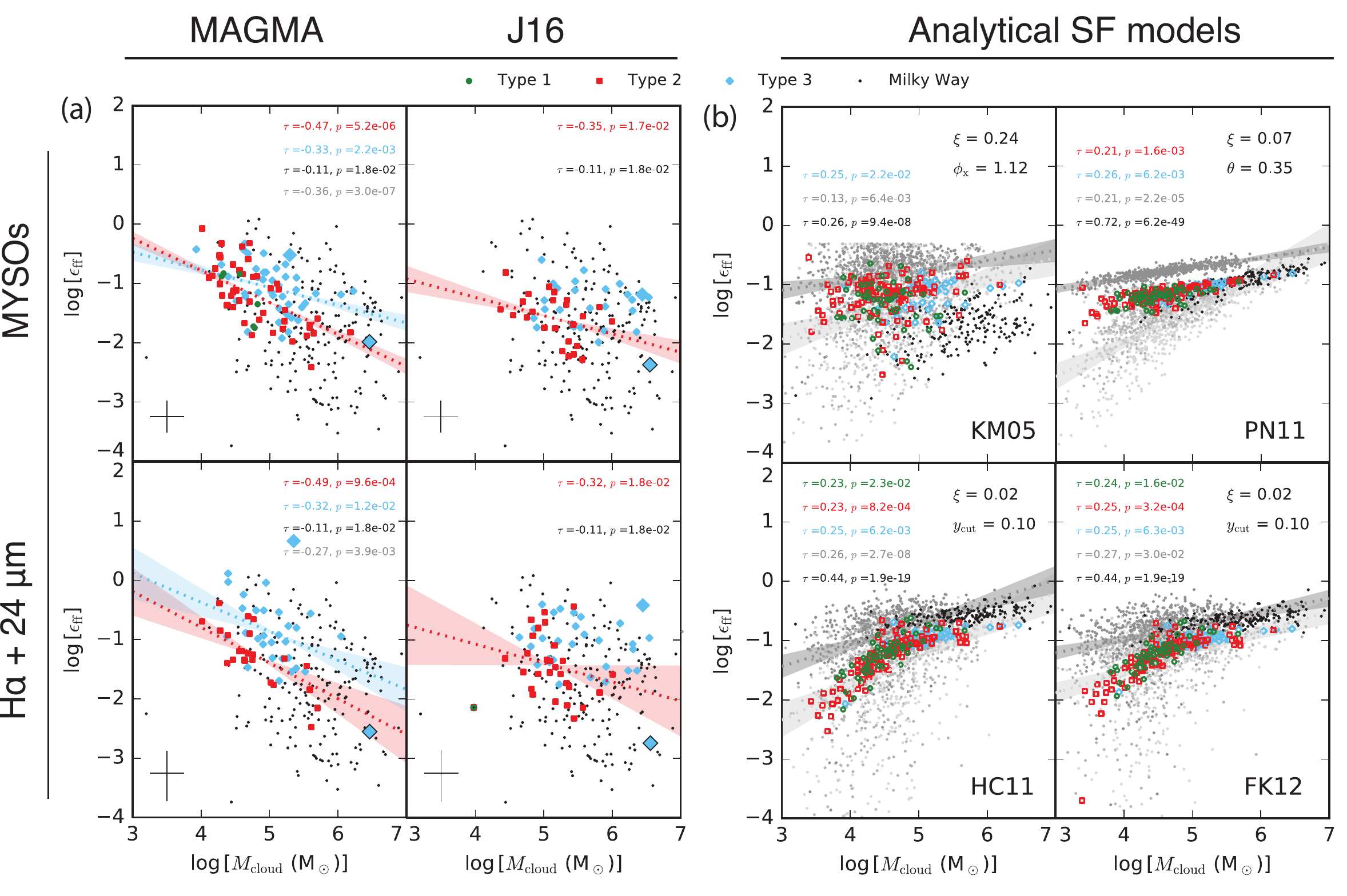} 
\caption{{\bf (a)}: Same as Fig. \ref{fig:sfe}b, but now for the star formation efficiency per free-fall time ($\epsilon_\m{ff}$) versus cloud mass ($M_\m{cloud}$). Also plotted here is the Galactic sample of \citet{lee_2016}. {\bf (b)}: $\epsilon_\m{ff}$ as predicted by analytical turbulence-regulated star formation models of \citet{krumholz_2005} (KM05; upperleft), \citet{padoan_2011} (PN11; upper right), \citet{hennebelle_2011} (HC11, lower left), and the \citet{hennebelle_2011} model extended with the effects of magnetic fields (\citealt{federrath_2012}; FK12). Shown are predictions for Type 1, 2, and 3 clouds assuming $b$ = 0.4 and $\beta$ = 0 for KM05, PN11, and HC11, while $\beta$ = 1.8 for the FK12 model (see text). The Milky Way clouds are shown with $b$ = 0.4 and $\beta$ = 0. Realizations of the model-predicted $\epsilon_\m{ff}$ for the LMC clouds are obtained by propagating all observational uncertainties(Tab. \ref{tab:model}) using solenoidal, mixed, and compressive forcing {\em (increasingly darker tones of gray)}. Fits are shown to the solenoidal and compressive realizations. Note that the KM05 model reaches an asymptotic value, mainly with full compressive forcing ($b$ = 1), since the error function inherent to the SFR model reaches a maximum (Tab. \ref{tab:model}).}
\label{fig:sfe_ff}
\end{figure*}

\subsubsection{Observations of $\epsilon_\m{ff}$}\label{sec:eff_obs}
In Fig. \ref{fig:sfe_ff}a, we plot $\epsilon_\m{ff}$ (= $\epsilon'$\,$\times$\,$\tau_\m{ff}$) versus $M_\m{cloud}$ for both the LMC and the Milky Way \citep{lee_2016}. While the large scatter in $\epsilon_\m{ff}$ on individual star formation region scales has been noted before \citep{mooney_1988, vutisalchavakul_2016,lee_2016}, the LMC data provides insight into the time evolution of $\epsilon_\m{ff}$. Indeed, a large part of the scatter in $\epsilon_\m{ff}$ stems from the systematic offset of $\epsilon_\m{ff}$ between Type 1, Type 2, and Type 3 clouds (Tab. \ref{tab:eff}), most notably for $\epsilon_\m{ff,H\alpha}$. As in Sec. \ref{sec:rates}, we fit the different subsets of cloud Types, which in all cases results in a negative slope $\alpha_\m{\epsilon_\m{ff}}$ (Tab. \ref{tab:eff}). 

The decline of $\epsilon_\m{ff}$ with $M_\m{cloud}$ (similar to that observed for $\epsilon'$; Fig. \ref{fig:sfe}) is also present in the Galactic samples of \citet{lee_2016}, \citet{murray_2011}, and \citet{vutisalchavakul_2016}. \citet{murray_2011} argues that this trend is largely explained through an observational selection effect introduced by a lower luminosity limit (shaded areas in Fig. \ref{fig:sfe}), and because $M_\m{cloud}$ appears on both axes. However, while an uncertainty in $M_\m{cloud}$ would scatter sources along a line with slope -1, this can not reconcile a {\em systematic} decline of $\epsilon_\m{ff}$ with $M_\m{cloud}$. Instead, we show that negative slopes derived for  $\epsilon'$ and $\epsilon_\m{ff}$ with $M_\m{cloud}$ originate from uncorrelated or sublinear correlated SFRs with $M_\m{cloud}$ (Sec. \ref{sec:rates} \& Tab. \ref{tab:sfr}). Combining the results of the LMC and the Milky Way, it becomes apparent that these relations are inherent to the star forming properties of individual GMCs when {\em total} cloud masses are measured through $^{12}$CO(1-0), dust, or $^{13}$CO(1-0).

Disruption of GMCs through stellar feedback may contribute to the observed scatter in $\epsilon_\m{ff}$. In other words, Type 3 clouds may have lost a significant amount of mass through destructive feedback from their associated clusters, thereby systematically altering $\epsilon_\m{ff}$ \citep[for a discussion, see][]{krumholz_2014}. The average age of massive star clusters powering SFR$_\m{H\alpha}$ is $\sim$\,4 Myr \citep{murray_2011}. Observations \citep{ochsendorf_2015b} and simulations \citep{dale_2014} of feedback on GMCs reveal typical photo-evaporation rates of $\sim$10$^{-2}$ M$_{\odot}$ yr$^{-1}$, implying that only $\sim$5\% of a 6\,$\times$\,10$^5$ M$_\m{\odot}$ GMC (average Type 3 cloud mass; \citealt{kawamura_2009}) is lost in $\sim$\,4 Myr. In fact, Type 3 clouds are slightly more massive than their Type 2 counterparts \citep{kawamura_2009}. Thus, it seems unlikely that GMC disruption alone can be responsible for the systematic increase in $\epsilon_\m{ff,MYSO}$ and $\epsilon_\m{ff,H\alpha}$ between Type 1, Type 2, and Type 3 clouds (Tab. \ref{tab:eff}). The timescale involved in GMC disruption may potentially be estimated through the uncertainty principle proposed by \citet{kruijssen_2014}, however, this method does not distinguish between GMC displacement/redistribution of star formation (e.g., \HII\ region expansion) and actual GMC disruption.

The large scatter in $\epsilon_\m{ff}$ may instead be driven by an increasing star formation rate (i.e., accelerating star formation) per free-fall time $\tau_\m{ff}$ in more evolved GMCs. Similar patterns have been observed in nearby clusters \citep{palla_2000} and M17 \citep{povich_2016}, and have been explored theoretically \citep{zamora-aviles_2012}. While it is clear that the star formation {\em rate} increases within Type 2 and Type 3 clouds (Sec. \label{sec:rates}), there is no significant change of $\tau_\m{ff}$ between cloud Types (Tab. \ref{tab:sfr}), driving $\epsilon_\m{ff}$ upward. Note, however, that $\tau_\m{ff}$ is calculated from a single average density. If the dense gas forming massive stars constitutes a small fraction of the total GMC mass, this will not be reflected in our estimation of $\tau_\m{ff}$.  The LMC and Milky Way data may therefore imply that the massive star forming rates and efficiencies of GMCs are not regulated by its global properties, but instead occurs only in specific, small volumes of GMCs \citep{evans_2014,ochsendorf_2016,vutisalchavakul_2016}. 

Finally, we note that $\langle\epsilon_\m{ff,MYSO}\rangle$ and $\langle\epsilon_\m{ff,H\alpha}\rangle$ for the LMC clouds (0.10 - 0.25; Tab. \ref{tab:eff}) are significantly higher than in Galactic clouds as measured by free-free flux ($\sim$\,0.03; \citealt{lee_2016}). However, given the similarity of $\epsilon_\m{ff}$ of GMCs at mass 4.0\,$\textless$\,log($M_\m{cloud}$)\,$\textless$\,5.5, this can largely be explained by a dearth of GMCs in the LMC at high $M_\m{cloud}$, which tend to have lower $\langle\epsilon_\m{ff}\rangle$.
  
\subsubsection{Predictions of $\epsilon_\m{ff}$ with analytical star formation models}\label{sec:eff_theory}

In this section, we compare the observations of the LMC and the Milky Way with analytical star formation models from \citet{krumholz_2005} (KM05), \citet{padoan_2011} (PN11), and \citet{hennebelle_2011} (HC11). In addition, we consider the \citet{hennebelle_2011} model extended to include magnetic fields by \citet{federrath_2012} (FK12). For a detailed explanation and comparison between these models, we refer the reader to \citet{federrath_2012} and \citet{padoan_2014}. 

\begin{table*}
\centering
\caption{Analytical star formation models, parameters, and observables}
\begin{tabular}{l|c|c}\hline \hline
Model & $x_\m{crit}$ & $\epsilon_\m{ff}$ \\ \hline \hline
KM05 & ($\pi^2$/45)$\phi_\m{x}^2$$\alpha_\m{vir}$$\mathcal{M}^2$ & $\xi$ $\left\{1 + \m{erf}[(\sigma_\m{p}^2 - 2\m{ln}\, x_\m{crit}) / (8\sigma_\m{p}^2)^{1/2}]\right\}$ \\
PN11 & (0.067)$\theta^{-2}$$\alpha_\m{vir}$$\mathcal{M}^2$$f(\beta)$ & $\xi$ $\left\{1 + \m{erf}[(\sigma_\m{p}^2 - 2\m{ln}\, x_\m{crit}) / (8\sigma_\m{p}^2)^{1/2}]\right\}$$e^{1/2 x_\m{crit}}$ \\
HC11 & ($\pi^2$/5)$y_\m{cut}^{-2}$$\alpha_\m{vir}$$\mathcal{M}^{-2}$ & $\xi$ $\left\{1 + \m{erf}[(\sigma_\m{p}^2 - \m{ln}\, x_\m{crit}) / (2\sigma_\m{p}^2)^{1/2}]\right\}$$e^{3\sigma_\m{p}^2/8}$ \\ 
FK12 & ($\pi^2$/5)$y_\m{cut}^{-2}$$\alpha_\m{vir}$$\mathcal{M}^{-2}$$(1 + \beta^{-1})^{-1}$ & $\xi$ $\left\{1 + \m{erf}[(\sigma_\m{p}^2 - \m{ln}\, x_\m{crit}) / (2\sigma_\m{p}^2)^{1/2}]\right\}$$e^{3\sigma_\m{p}^2/8}$ \\  \hline \hline

Parameters & Value & Reference \\ \hline \hline
$\phi_\m{x}$ & 1.19 & \citet{krumholz_2005} \\
$y_\m{cut}$ & 0.1 & \citet{hennebelle_2011} \\
$\theta$ & 0.35 & \citet{padoan_2011} \\
$b$ & 0.33, 0.4, 1.0 & \citet[]{federrath_2012} \\ 
$\beta$ & 0.04, 0.20, 1.8, 3.6 &  \citet[]{federrath_2012} \\ \hline \hline

Observables & Uncertainty (dex) & Reference \\  \hline \hline
$v_\m{CO}$ & 0.08 & \citet{wong_2011} \\
$c_\m{s}$ & 0.20 & \citet{roman-duval_2010} \\
$R_\m{cloud}$ & 0.11 & \citet{wong_2011} \\
$M_\m{cloud}$ & 0.30 & \citet{bolatto_2013} \\ \hline
\end{tabular}
\tablecomments{The analytical descriptions of the star formation models (formulations similar to that in \citealt{federrath_2012} and \citealt{padoan_2014}), the parameters used in the model realizations, and the observables with their estimated uncertainties. The function $f$($\beta$) is given in \citet{padoan_2011}. The turbulent forcing parameter $b$ and plasma $\beta$ are chosen to be consistent with those used in \citet{federrath_2012}. The uncertainties on the observables are assumed to be normally distributed.}
\label{tab:model}
\end{table*}

The aforementioned models rely on the PDF of gas density, often assumed to be log-normal in turbulent media \citep{vazquez-semadeni_1994,padoan_1997,burkhart_2012}. The width of the density PDF is determined by $\sigma_\m{p}$ = ln(1 + $b^2$$\mathcal{M}$), where $b$ is the turbulence forcing parameter ($b$ = 0.33, 0.4, and 1 for purely solenoidal, mixed, or purely compressive forcing, respectively; \citealt{federrath_2012}), and $\mathcal{M}$ is the internal Mach number ($\mathcal{M}$ = $v_\m{CO}$/$c_\m{s}$; where $c_\m{s}$ is the isothermal speed of sound). The star formation rate per free-fall time, $\epsilon_\m{ff}$, is then determined by estimating the gas mass above a given density threshold, $x_\m{crit}$. The exact definition of $x_\m{crit}$ differs between individual models (Tab. \ref{tab:model}), but is dependent on the virial parameter $\alpha_\m{vir}$ = 5$v_\m{CO}$$R_\m{cloud}$/$GM_\m{cloud}$ (where $R_\m{cloud}$ and $v_\m{CO}$ are the radius and CO 1D linewidth), $\mathcal{M}$, $b$, and the plasma $\beta$ = 2$\mathcal{M}$/$\mathcal{M_\m{A}}$, where $\mathcal{M_\m{A}}$ is the Alfv\'{e}nic Mach number relating to the strength of the magnetic field.

We measure the second velocity moment (Wong et al., in prep) within our cloud footprints (Sec. \ref{sec:cloud}) to determine $v_\m{CO}$. We then perform model realizations by propagating uncertainties of all observables $v_\m{co}$, $c_\m{s}$, $R_\m{cloud}$, and $M_\m{cloud}$, while randomly choosing between specified values of the turbulent forcing parameter $b$ and the plasma $\beta$ (Tab. \ref{tab:model}). For $c_\m{s}$, we opted to choose a distribution with mean value of 0.2 km\,s$^{-1}$ (for a gas temperature of 15 K) with an 1$\sigma$ uncertainty of $\pm$0.2 dex to account for differences in UV field or cosmic ray environments. This spread is based on the CO excitation temperatures of Galactic GMCs \citep{roman-duval_2010}. As excitation temperature may differ from the actual gas temperature at low densities due to sub-thermal excitation, we only use the spread of excitation temperatures derived in their sample to estimate the uncertainty on the cloud temperatures and thereby $c_\m{s}$ (Tab. \ref{tab:model}).

\begin{table*}
\centering
\caption{Star formation efficiencies per free fall time: observations versus theory}
\begin{tabular}{l|c|c|c|c||c|c|c|c||c}\hline

& \multicolumn{8}{c}{\bf Observations} \\ \hline
& \multicolumn{4}{c||}{\bf MAGMA} & \multicolumn{4}{c||}{\bf J16} & {\bf MW}  \\ \hline
& Type 1 & Type 2 & Type 3 & All  & Type 1 & Type 2 & Type 3 & All & \\  \hline
$\langle\epsilon_\m{ff,MYSO}\rangle$  &  0.08 & 0.12 & 0.12 & 0.12 &  & 0.04 & 0.06 & 0.05 &  \\
$\langle\epsilon_\m{ff,H\alpha}\rangle$ &  & 0.09 & 0.37 & 0.25 & & 0.07 & 0.12 & 0.10 & 0.03 \\
$\langle\epsilon_\m{ff,H\alpha}\rangle$$^{(*)}$ &  & 0.09 & 0.24 & 0.17 & & 0.07 & 0.12 & 0.10 & 0.03 \\

$\alpha_\m{\epsilon_\m{ff,MYSO}}$ & & -0.54$\pm$0.07 & -0.31$\pm$0.07 & -0.37$\pm$0.05 &  & -0.30$\pm$0.10 & & \\

$\alpha_\m{\epsilon_\m{ff,H\alpha}}$ &  & -0.56$\pm$0.25 & -0.53$\pm$0.21 & -0.38$\pm$0.14 &  & -0.30$\pm$0.29 & & & -0.25 \\
$\alpha_\m{\epsilon_\m{ff,H\alpha}}$$^{(*)}$ &  & -0.59$\pm$0.27 & -0.39$\pm$0.20 & -0.30$\pm$0.15 &  & -0.28$\pm$0.31 & & & -0.25 \\

$\sigma_\m{log\,\epsilon_\m{ff,MYSO}}$ & & 0.38 & 0.37 & 0.41 & & & 0.35 & & \\ 
$\sigma_\m{log\,\epsilon_\m{ff,H\alpha}}$ & & 0.36 & 0.57 & 0.58 & & 0.47 & & & 0.91 \\ 
$\sigma_\m{log\,\epsilon_\m{ff,H\alpha}}$$^{(*)}$ & & 0.36 & 0.44 & 0.49 & & 0.47 & & & 0.91 \\ \hline

& \multicolumn{8}{c}{\bf Analytical models} \\ \hline
& \multicolumn{4}{c||}{\bf MAGMA} & \multicolumn{4}{c||}{\bf Turbulent forcing parameters} & {\bf MW}$^{(**)}$  \\ \hline
& Type 1 & Type 2 & Type 3 & All  & $b$ = 0.33 & $b$ = 0.40 &  $b$ = 1.00 & All & \\  \hline
$\langle\epsilon_\m{ff,KM05}\rangle$ &  0.06 & 0.07 & 0.07 & 0.07 & 0.04 & 0.05 & 0.14 & 0.07 & 0.04 \\ 
$\langle\epsilon_\m{ff,PN11}\rangle$ & 0.05 & 0.05 & 0.06 & 0.05 & 0.02 & 0.04 & 0.16 & 0.05 & 0.12 \\
$\langle\epsilon_\m{ff,HC11}\rangle$ & 0.05 & 0.05 & 0.09 & 0.06 & 0.03 & 0.05 & 0.14 & 0.06 & 0.27 \\ 
$\langle\epsilon_\m{ff,FK12}\rangle$ & 0.05 & 0.05 & 0.07  & 0.05 & 0.05 & 0.05 & 0.07 & 0.08 & 0.23 \\ \hline

$\alpha_\m{\epsilon_\m{ff,KM05}}$ & 0.34$\pm$0.18 & 0.29$\pm$0.06 & 0.29$\pm$0.10&  0.25$\pm$0.07 & 0.27$\pm$0.11 & 0.29$\pm$0.13 & 0.22$\pm$0.12 & 0.26$\pm$0.06 & 0.21 \\
$\alpha_\m{\epsilon_\m{ff,PN11}}$ & 0.40$\pm$0.19 & 0.43$\pm$0.11 & 0.37$\pm$0.10 &  0.41$\pm$0.04 & 0.61$\pm$0.11 & 0.49$\pm$0.09 & 0.18$\pm$0.05 &  0.42$\pm$0.08 & 0.30 \\
$\alpha_\m{\epsilon_\m{ff,HC11}}$ & 0.52$\pm$0.25 & 0.47$\pm$0.09 & 0.34$\pm$0.10 &  0.40$\pm$0.05 & 0.58$\pm$0.08 & 0.49$\pm$0.11 & 0.38$\pm$0.08 &  0.44$\pm$0.06 & 0.15 \\ 
$\alpha_\m{\epsilon_\m{ff,FK12}}$ & 0.51$\pm$0.23 & 0.40$\pm$0.08 & 0.30$\pm$0.09&  0.43$\pm$0.05 & 0.36$\pm$0.10 & 0.38$\pm$0.12 & 0.27$\pm$0.09 & 0.32$\pm$0.05 & 0.15 \\ \hline
$\sigma_\m{log\,\epsilon_\m{ff,KM05}}$ & 0.61 & 0.55 & 0.49 & 0.55 & 0.59 & 0.52 & 0.33 & 0.56 & 0.28 \\ 
$\sigma_\m{log\,\epsilon_\m{ff,PN11}}$ & 0.47 & 0.45 & 0.42 & 0.47 & 0.38 & 0.30 & 0.09 & 0.48 & 0.10 \\ 
$\sigma_\m{log\,\epsilon_\m{ff,HC11}}$ & 0.74 & 0.64 & 0.52 & 0.65 & 0.64 & 0.59 & 0.48 & 0.65 & 0.19 \\ 
$\sigma_\m{log\,\epsilon_\m{ff,FK13}}$ & 0.63 & 0.65 & 0.56 & 0.62 & 0.62 & 0.54 & 0.33 & 0.51 & 0.17 \\ \hline

\end{tabular}
\tablecomments{Listed are: the average star formation rate per free-fall time measured with MYSOs ($\langle\epsilon_\m{ff,MYSO}\rangle$) and H$\alpha$ ($\langle\epsilon_\m{ff,H\alpha}\rangle$); slope indices $\alpha_\m{ff}$; and the scatter around the fitted regression ($\sigma_\m{log\,\epsilon_\m{ff}}$; in dex). Then the same parameters are listed for the \citet{krumholz_2015} (KM05), \citet{padoan_2011} (PN11) \citet{hennebelle_2011} (HC11), and the \citet{federrath_2012} (FK12) turbulence-regulated star formation models (see text). $^{(*)}$: These values refer to the fits by excluding the 30 Doradus and SMR region (see text). $^{(**)}$: Observational uncertainties for several of the values in Tab. \ref{tab:model} are not given in \citet{lee_2016} and are thus not propagated. In addition, for consistency between the LMC and the Galaxy, these values are calculated using the normalization constant $\xi$ of the LMC and $b$ = 0.4 (see text and Fig. \ref{fig:sfe_ff}).}
\label{tab:eff}
\end{table*}

To facilitate direct comparisons between observations and theory, we must choose for each model the unique parameter that prescribes the criterium for gravitational collapse (i.e., the model `fudge factor'; \citealt{federrath_2012}). We adopted the author-preferred values, i.e., $\phi_\m{x}$ = 1.12 (KM05), $\theta$ = 0.35 (PN11), and $y_\m{cut}$ = 0.1 (HC11 \& FK12). In addition, the normalization constant $\xi$, which relates to the core efficiency and characteristic timescale over which gas becomes gravitationally unstable, is chosen such that the median value of the model realization matches that of the LMC sample. The results are shown in Fig. \ref{fig:sfe_ff}b. 

The most striking discrepancy between model and observations is the derived slope $\alpha_\m{\epsilon_\m{ff}}$. None of the analytical star formation models is able to reproduce the decline of $\epsilon_\m{ff}$ with $M_\m{cloud}$ that is seen in the observations of the LMC and the Milky Way. The increase of $\epsilon_\m{ff}$ with $M_\m{cloud}$ in the models likely results from the GMC size-linewidth relation \citep{larson_1981,bolatto_2008,heyer_2009,roman-duval_2010,wong_2011}. In this case, larger clouds will inevitably have higher $\mathcal{M}$, where theory predicts higher $\epsilon_\m{ff}$ as stronger and denser compression leads to higher SFRs. We have verified that the discrepancy between model and observations is independent of any choice/combination of fudge factor, normalization constant, and level of observational uncertainties.

Milky Way GMCs have larger $v_\m{CO}$ and $\alpha_\m{vir}$, which tend to decrease $\epsilon_\m{ff}$, but this is counteracted by a simultaneous increase in $\mathcal{M}$, elevating $\epsilon_\m{ff}$. Consequently, at the same $\xi$ and model fudge factors, the LMC and Milky Way clouds populate a continuous parameter space in Fig. \ref{fig:sfe_ff}b in the PN11, HC11, and FK12 models. However, in the KM05 model the LMC and Milky Way clouds significantly differ in $\epsilon_\m{ff}$ since the KM05 model is relatively insensitive to an increase in Mach number \citep{krumholz_2005,federrath_2012}. 

None of the models predict significant changes in $\epsilon_\m{ff}$ between cloud Types, which is likely because $v_\m{CO}$ is independent of star formation rate (Tab. \ref{tab:sfr}) as turbulent driving and dissipation scale together \citep{kim_2011}. Nonetheless, the observed offset between cloud Types can be reproduced if the cloud Types experience different turbulent forcing, i.e., solenoidal, mixed, or compressive (Fig. \ref{fig:sfe_ff}). Indeed, Type 3 clouds contain nearby stellar clusters and may encountered a higher rate of (nearby) stellar feedback, which may inject substantial compressive forcing locally (i.e., higher $b$). Nonetheless, any cloud compression must be acting on a relative small mass fraction since $\tilde{\tau_\m{ff}}$ does not differ significantly between Type 1, 2, and 3 clouds. 

The model realizations reveal a large scatter similar to the observations when all observational uncertainties are propagated (see Tab. \ref{tab:eff}), which was not performed by \citet{lee_2016}. However, the model-predicted scatter is largely driven by lower mass clouds (Fig. \ref{fig:sfe_ff}), which is not reflected in the observations. This indicates that a different physical mechanism drives the scatter in the observations which is not captured in the analytical star formation models (See Sec. \ref{sec:discussion}).

\section{Discussion}\label{sec:discussion}

\subsection{Star formation properties on individual GMC scales}

In the LMC, the SFR of GMCs increases over multiple generations (i.e., SFR$_\m{MYSO}$ and SFR$_\m{H\alpha}$) across the evolutionary sequence of GMCs proposed by \citet{kawamura_2009}. Whereas SFR$_\m{H\alpha}$ is known to be affected by various systematics creating scatter and potential bias (Sec. \ref{sec:indirect}), direct counting of massive YSOs is not affected in this sense. The relative contribution of SFR$_\m{MYSO}$ and SFR$_\m{H\alpha}$ varies between cloud Types, but both tracers show good agreement for the entire ensemble of GMCs (Tab. \ref{tab:sfr}), albeit with scatter that is expected for stochastic sampling of the high-end tail of the IMF at these SFRs  (Tab. \ref{fig:sfr_tracers}; \citealt{dasilva_2012,krumholz_2014}). Comparison of SFRs from indirect star formation tracers with (massive) YSO counting on large scales remain complicated in the Milky Way because of confusion and distance ambiguities. Such studies are currently limited to the Magellanic system, but will soon be possible in nearby galaxies with the upcoming James Webb Space Telescope.

The increase of SFR between cloud Types drives the observed scatter of $\epsilon'$ and $\epsilon_\m{ff}$ with $M_\m{cloud}$, and it is very likely that the observed scatter in the Milky Way originates from the same physical mechanism \citep{mooney_1988,mead_1990,murray_2010,lee_2016,vutisalchavakul_2016}. In addition, we have determined that both $\epsilon'$ and $\epsilon_\m{ff}$ decline with $M_\m{cloud}$ (Figs. \ref{fig:sfe}b and \ref{fig:sfe_ff}a), as SFRs are not linearly correlated with total $M_\m{cloud}$ on individual cloud scales in the LMC and the Galaxy (Fig. \ref{fig:sfe}a and Tab. \ref{tab:sfr}). Massive clouds appear to have larger diffuse (i.e., non star forming) envelopes, which potentially affects the global star formation potential of GMCs (Fig. \ref{fig:gmc_comp}). One implication of this result is a higher molecular depletion time $\tau_\m{dep}$ with increasing GMC mass. This may be related to trends observed recently on kpc and galactic scales by \citet{leroy_2013} and \citet{bigiel_2016}, who find higher $\tau_\m{dep}$ in regions of high surface density. 

\subsection{Observations versus theory}

We have tested the turbulence-regulated analytical star formation models of \citet{krumholz_2005}, \citet{padoan_2011}, \citet{hennebelle_2011}, \citet{federrath_2012} and shown that these are unable to reproduce the decline of $\epsilon_\m{ff}$ with increasing $M_\m{cloud}$ (Fig. \ref{fig:sfe_ff}). This discrepancy likely originates from the GMC size-linewidth relation, resulting in higher Mach numbers and higher $\epsilon_\m{ff}$ in all considered models (Sec. \ref{sec:eff_theory}). In this regard, several authors have questioned the existence of the size-linewidth relation \citep{vazquez-semadeni_1997,ballesteros-paredes_2002, ballesteros-paredes_2012} because of limited dynamic range of observations. These challenges were in its turn challenged by \citet{lombardi_2010} using near-infrared extinction mapping, a technique that can probe large dynamic ranges. Regardless of the method used, an intrinsic column density threshold defining molecular clouds may introduce a bias as this will affect clouds of varying sizes differently. Further research to disentangle the effects of GMC definition and its impact on derived star formation properties is necessary in this sense.

In relation to the above, our results emphasize the importance of a consistent definition of GMCs. Comparison of SFR, $\epsilon'$, and $\epsilon_\m{ff}$ between two LMC cloud sets (MAGMA and J16) reveal significant difference in the absolute values of these quantities. Unfortunately, in many cases the exact definition of GMCs will be dictated by the sensitivity, resolution, and the cloud decomposition of observations, as well as the used tracer of molecular material. Once resolved maps of nearby galaxies become readily available (e.g., this work, \citealt{bigiel_2016}), this may complicate comparison of star formation rates and efficiencies of nearby clouds \citep{evans_2009, lada_2010, heiderman_2010} with those in galaxies at different metallicity and/or surface density \citep{hughes_2013}. Nonetheless, the main results of this work (i.e., the systematic offset between cloud Types and the decline of $\epsilon'$ and $\epsilon_\m{ff}$ with cloud mass) are unchanged irregardless of the used cloud map.

While the absolute scatter of the observations can be reproduced with the star formation models, the model-predicted scatter is largely driven by lower mass clouds (Fig. \ref{fig:sfe_ff}), which is not in agreement with the observations. Instead, we have shown that the increasing SFR as a function of time can perfectly drive the scatter (Secs. \ref{sec:rates} and \ref{sec:eff}). Our results may therefore confirm that SFR is a time-variable quantity, which appears to increase as soon as stellar clusters emerge and start to disrupt the parent GMCs \citep{murray_2011, lee_2016}. 

While the \citet{krumholz_2005}, \citet{padoan_2011}, \cite{hennebelle_2011}, and \citet{federrath_2012} models naturally explain observed core mass functions at masses of $\lesssim$\,10 M$_\odot$ \citep{andre_2010}, suggesting that large-scale turbulent properties of GMCs are important in low-to-intermediate mass star formation \citep{padoan_2014}, it has been unclear if this extends to the formation of massive stars \citep{offner_2014}. Here, we have shown that these analytical models, idealizing global turbulence levels, cloud densities, and assuming a stationary SFR do not reproduce observations from modern large datasets tracing massive star formation on galaxy-wide scales (this work, \citealt{lee_2016}). The discrepancy may originate from the assumption of a stationary SFR in the models, but may also reflect the fact that {\em global} turbulent properties are irrelevant to the formation of massive stars. 

\subsection{What sets the massive star formation rate of GMCs?}

The large scatter in SFR and SFE at individual cloud scales arises when {\em total} GMC masses are considered (this work, \citealt{mooney_1988,mead_1990,murray_2010,lee_2016,vutisalchavakul_2016}). In contrast, SFR shows a roughly linear relation with dense gas with significantly less scatter \citep{wu_2005,lada_2010,heiderman_2010,evans_2014,vutisalchavakul_2016}. This illustrates the inability of $^{12}$CO(1-0), $^{13}$CO(1-0), or dust-based measures of molecular gas to probe the actual star-forming component of the parent GMC \citep[e.g.,][]{gao_2004}: the observed scatter in SFE then reflects a varying fraction of dense gas in GMCs \citep{lada_2012,krumholz_2012}. 

Following the above, the observed decline in SFE and $\epsilon_\m{ff}$ with $M_\m{cloud}$ then simply reflects that the amount of gas participating in the formation of massive stars does not scale with total cloud mass. This is substantiated by the observation that GMCs mostly grow by adding diffuse (`CO-dark') gas (Fig. \ref{fig:gmc_comp}). Indeed, massive star formation appears to occupy only a small volume of its parent GMC \citep{ochsendorf_2016}, configurations that are readily recognized in nearby massive star forming regions such as Orion \citep{bally_1987} and M17 \citep{povich_2007}. In this regard, it has been questioned if the Kennicutt-Schmidt Law \citep{schmidt_1959,kennicutt_1998} is a result of an actual underlying star formation law, or simply a means of counting clouds \citep{lada_2012}. In other words, by averaging over kiloparsec scales, one mainly traces gas which may be irrelevant to the process of massive star formation.

It is clear that the dense gas fraction is crucial to the massive star forming properties of GMCs. However, the physical mechanisms regulating the dense gas is less clear. In this sense, the increase of SFRs between Type 1, Type 2, and Type 3 clouds and the decline of $\epsilon'$ and $\epsilon_\m{ff}$ with increasing cloud mass (Figs. \ref{fig:sfe}, \ref{fig:boxplot}, and \ref{fig:sfe_ff}) may show that stellar feedback injected {\em locally} is of key importance in shaping GMCs by controlling the dense gas fraction, and thereby the massive star formation rate. In other words, the first effect of massive star feedback is `positive', increasing the local star formation rate \citep{elmegreen_1977}. Only later does feedback seem to be negative \citep{hopkins_2011}; understanding this transition will be important for a thorough understanding of the massive star forming process. 

\section{Conclusions}\label{sec:conclusions}
We have studied massive star formation on individual cloud scales across the LMC, complemented this with Milky Way observations, and contrasted our results with widely-used analytical star formation models. Our main conclusions are as follows:

\begin{enumerate}

\item Star formation rates of GMCs in the LMC {\em increase} along the evolutionary sequence proposed by \citet{kawamura_2009}. This increase in SFR is sustained over several generations of massive stars, as both SFR$_\m{H\alpha}$ and SFR$_\m{MYSO}$ rise, which trace independent, separate generations of massive stars.

\item Both the star formation `efficiency' ($\epsilon'$ = SFR/$M_\m{cloud}$) and star formation efficiency per free fall time ($\epsilon_\m{ff}$ = $\epsilon'$\,$\times$\,$\tau_\m{ff}$) {\em decrease} as a function of total cloud mass in the LMC and Milky Way. This implies higher depletion times for larger clouds, possibly because GMCs mostly grow by adding diffuse (`CO-dark') gas. 

\item Analytical `turbulence-regulated' star formation models do {\em not} reproduce recent datasets tracing massive star formation on individual cloud scales in the LMC and Milky Way.

\end{enumerate}

\acknowledgements We thank the referee for a constructive report, the MAGMA team for permission to use the DR3 products in advance of publication, and Katie Jameson for providing the dust-based molecular hydrogen map. BBO is supported through NASA ADAP grant NNX15AF17G.

\appendix

Here we provide the results of our cloud decomposition and SFR measurements in four machine-readable tables: MAGMA clouds and MYSOs (Tab. \ref{tab:a1}), MAGMA clouds and Ha\,+\,24\,$\mu$m (Tab. \ref{tab:a2}), J16 clouds and MYSOs (Tab. \ref{tab:a3}), and J16 clouds and Ha\,+\,24\,$\mu$m (Tab. \ref{tab:a4}).

In Tables \ref{tab:a1} and \ref{tab:a3}, we report {\em all} GMCs (including those without detected MYSOs, i.e., SFR$_\m{MYSO}$ = 0) that were used to calculate the GMC parameters in Table \ref{tab:sfr}. However, the SFR parameters (SFR, $\epsilon'$, and $\epsilon_\m{ff}$) were calculated by only taking into account clouds with detected star formation (SFR$_\m{MYSO}$\,$\textgreater$\,0). Tables \ref{tab:a2} and \ref{tab:a4} only report those clouds with SFR$_\m{H\alpha}$\,$\textgreater$\,0.

In many cases, the H$\alpha$ and 24\,$\mu$m emission is extended and overlaps multiple GMCs (Sec. \ref{sec:match}). In these cases, cloud masses are added, radii are calculated assuming spherical symmetry, and central coordinates are estimated by weighting all individual GMC components by intensity (i.e., mass).

\begin{table*}
\centering
\caption{MAGMA cloud decomposition with MYSO SFR parameters}
\begin{tabular}{l|c|c|c|c|c|c|c|c|c|c}\hline \hline
Cloud No. & RA & DEC & Type & $M_\m{cloud}$ & $R_\m{cloud}$ & SFR$_\m{MYSO}$ & $\epsilon'_\m{MYSO}$ & $\epsilon_\m{ff,MYSO}$ & $\tau_\m{ff}$ & $v_\m{CO}$ \\
 & (deg) & (deg) & & ($M_\odot$) & (pc) & ($M_\odot$\,Myr$^{-1}$) & (Myr$^{-1}$) & & (Myr) & (km\,s$^{-1}$) \\  \hline
1\_mag\_myso&80.753&-68.322&1&6.41e+03&1.05e+01&0.00e+00&0.00e+00&0.00e+00&7.01e+00&1.19e+00\\
2\_mag\_myso&77.186&-69.071&1&7.51e+03&1.10e+01&0.00e+00&0.00e+00&0.00e+00&6.96e+00&1.22e+00\\
3\_mag\_myso&80.464&-69.810&1&7.86e+03&1.14e+01&0.00e+00&0.00e+00&0.00e+00&7.20e+00&7.54e-01\\
4\_mag\_myso&86.528&-71.120&1&7.93e+03&1.26e+01&0.00e+00&0.00e+00&0.00e+00&8.29e+00&1.17e+00\\
5\_mag\_myso&84.084&-71.408&1&8.63e+03&1.19e+01&0.00e+00&0.00e+00&0.00e+00&7.30e+00&9.17e-01\\
6\_mag\_myso&78.052&-67.755&1&1.05e+04&1.35e+01&0.00e+00&0.00e+00&0.00e+00&7.99e+00&1.37e+00\\
7\_mag\_myso&83.815&-68.223&1&1.12e+04&1.41e+01&0.00e+00&0.00e+00&0.00e+00&8.21e+00&7.94e-01\\
8\_mag\_myso&80.139&-68.552&1&1.29e+04&1.49e+01&0.00e+00&0.00e+00&0.00e+00&8.36e+00&1.03e+00\\
9\_mag\_myso&86.891&-68.139&1&1.43e+04&1.51e+01&0.00e+00&0.00e+00&0.00e+00&8.13e+00&9.84e-01\\
10\_mag\_myso&87.619&-68.442&1&1.55e+04&1.58e+01&0.00e+00&0.00e+00&0.00e+00&8.34e+00&1.11e+00\\ \hline
\end{tabular}
\tablecomments{Table 4 is published in its entirety in the machine-readable format. A portion is shown here for guidance regarding its form and content.}
\label{tab:a1}
\end{table*}

\begin{table*}
\centering
\caption{MAGMA cloud decomposition with H$\alpha$\,+\,24 $\mu$m SFR parameters}
\begin{tabular}{l|c|c|c|c|c|c|c|c|c|c}\hline \hline
Cloud No. & RA & DEC & Type & $M_\m{cloud}$ & $R_\m{cloud}$ & SFR$_\m{H\alpha}$ & $\epsilon'_\m{H\alpha}$ & $\epsilon_\m{ff,H\alpha}$ & $\tau_\m{ff}$ & $v_\m{CO}$ \\
 & (deg) & (deg) & & ($M_\odot$) & (pc) & ($M_\odot$\,Myr$^{-1}$) & (Myr$^{-1}$) & & (Myr) & (km\,s$^{-1}$) \\  \hline
1\_mag\_ha24&76.749&-70.720&2&1.03e+04&1.28e+01&2.83e+02&2.74e-02&2.03e-01&7.41e+00&7.92e-01\\
2\_mag\_ha24&86.141&-69.736&2&1.83e+04&1.87e+01&7.76e+02&4.23e-02&4.16e-01&9.83e+00&1.19e+00\\
3\_mag\_ha24&80.039&-66.873&2&1.89e+04&1.59e+01&2.51e+02&1.33e-02&1.01e-01&7.65e+00&9.16e-01\\
4\_mag\_ha24&81.396&-69.312&3&2.38e+04&1.79e+01&3.29e+02&1.38e-02&1.13e-01&8.14e+00&1.06e+00\\
5\_mag\_ha24&75.578&-69.055&2&2.40e+04&1.97e+01&8.69e+01&3.61e-03&3.36e-02&9.30e+00&1.62e+00\\
6\_mag\_ha24&80.695&-69.856&2&2.46e+04&1.97e+01&1.88e+02&7.65e-03&7.04e-02&9.20e+00&1.33e+00\\
7\_mag\_ha24&74.317&-68.435&3&2.48e+04&1.82e+01&3.55e+03&1.44e-01&1.17e+00&8.18e+00&1.53e+00\\
8\_mag\_ha24&78.470&-67.398&3&2.48e+04&1.89e+01&2.70e+03&1.09e-01&9.39e-01&8.62e+00&1.42e+00\\
9\_mag\_ha24&76.581&-70.184&2&2.93e+04&2.05e+01&1.49e+02&5.09e-03&4.55e-02&8.94e+00&9.85e-01\\
10\_mag\_ha24&83.179&-69.770&2&3.62e+04&2.02e+01&2.93e+02&8.08e-03&6.36e-02&7.88e+00&1.30e+00\\ \hline
\end{tabular}
\tablecomments{Table 5 is published in its entirety in the machine-readable format. A portion is shown here for guidance regarding its form and content.}
\label{tab:a2}
\end{table*}

\begin{table*}
\centering
\caption{J16 cloud decomposition with MYSO SFR parameters}
\begin{tabular}{l|c|c|c|c|c|c|c|c|c}\hline \hline
Cloud No. & RA & DEC & Type & $M_\m{cloud}$ & $R_\m{cloud}$ & SFR$_\m{MYSO}$ & $\epsilon'_\m{MYSO}$ & $\epsilon_\m{ff,MYSO}$ & $\tau_\m{ff}$ \\
 & (deg) & (deg) & & ($M_\odot$) & (pc) & ($M_\odot$\,Myr$^{-1}$) & (Myr$^{-1}$) & & (Myr) \\ \hline
1\_j16\_myso&85.961&-71.130&1&3.75e+04&2.30e+01&0.00e+00&0.00e+00&0.00e+00&9.39e+00\\
2\_j16\_myso&86.077&-71.457&1&5.73e+04&2.97e+01&0.00e+00&0.00e+00&0.00e+00&1.11e+01\\
3\_j16\_myso&86.894&-68.138&1&6.39e+04&3.16e+01&0.00e+00&0.00e+00&0.00e+00&1.16e+01\\
4\_j16\_myso&87.050&-70.623&1&6.93e+04&3.42e+01&0.00e+00&0.00e+00&0.00e+00&1.26e+01\\
5\_j16\_myso&71.878&-67.220&1&7.31e+04&3.16e+01&0.00e+00&0.00e+00&0.00e+00&1.08e+01\\
6\_j16\_myso&85.849&-70.203&1&1.07e+05&3.75e+01&0.00e+00&0.00e+00&0.00e+00&1.16e+01\\
7\_j16\_myso&77.452&-69.204&1&1.28e+05&4.19e+01&0.00e+00&0.00e+00&0.00e+00&1.25e+01\\
8\_j16\_myso&80.629&-68.349&1&1.49e+05&4.19e+01&0.00e+00&0.00e+00&0.00e+00&1.16e+01\\
9\_j16\_myso&76.977&-69.001&1&1.93e+05&4.84e+01&0.00e+00&0.00e+00&0.00e+00&1.27e+01\\
10\_j16\_myso&81.705&-68.637&2&2.36e+04&1.71e+01&1.13e+02&4.76e-03&3.62e-02&7.61e+00\\ \hline
\end{tabular}
\tablecomments{Table 6 is published in its entirety in the machine-readable format. A portion is shown here for guidance regarding its form and content.}
\label{tab:a3}
\end{table*}

\begin{table*}
\centering
\caption{J16 cloud decomposition with H$\alpha$\,+\,24 $\mu$m SFR parameters}
\begin{tabular}{l|c|c|c|c|c|c|c|c|c}\hline \hline
Cloud No. & RA & DEC & Type & $M_\m{cloud}$ & $R_\m{cloud}$ & SFR$_\m{H\alpha}$ & $\epsilon'_\m{H\alpha}$ & $\epsilon_\m{ff,H\alpha}$ & $\tau_\m{ff}$ \\
 & (deg) & (deg) & & ($M_\odot$) & (pc) & ($M_\odot$\,Myr$^{-1}$) & (Myr$^{-1}$) & & (Myr) \\ \hline
1\_j16\_ha24&76.331&-66.896&2&2.74e+04&1.71e+01&1.90e+02&6.93e-03&4.90e-02&7.07e+00\\
2\_j16\_ha24&79.366&-71.233&3&4.39e+04&2.30e+01&5.59e+02&1.27e-02&1.10e-01&8.67e+00\\
3\_j16\_ha24&78.087&-70.414&2&5.00e+04&2.76e+01&1.34e+02&2.68e-03&2.87e-02&1.07e+01\\
4\_j16\_ha24&77.683&-67.081&2&5.78e+04&2.97e+01&3.11e+02&5.39e-03&5.98e-02&1.11e+01\\
5\_j16\_ha24&86.305&-69.797&3&6.47e+04&2.76e+01&5.32e+02&8.22e-03&7.74e-02&9.42e+00\\
6\_j16\_ha24&81.815&-70.573&2&6.51e+04&2.97e+01&1.36e+03&2.09e-02&2.19e-01&1.05e+01\\
7\_j16\_ha24&83.896&-67.718&2&6.60e+04&3.25e+01&8.21e+01&1.24e-03&1.48e-02&1.19e+01\\
8\_j16\_ha24&83.982&-69.529&2&6.91e+04&2.54e+01&1.00e+03&1.45e-02&1.17e-01&8.04e+00\\
9\_j16\_ha24&75.561&-69.039&2&6.93e+04&2.76e+01&9.10e+01&1.31e-03&1.20e-02&9.10e+00\\
10\_j16\_ha24&73.207&-68.077&3&7.24e+04&3.16e+01&1.02e+03&1.41e-02&1.54e-01&1.09e+01\\ \hline
\end{tabular}
\tablecomments{Table 7 is published in its entirety in the machine-readable format. A portion is shown here for guidance regarding its form and content.}
\label{tab:a4}
\end{table*}



\end{document}